\documentclass[aps,prx,twocolumn,notitlepage,nofootinbib]{revtex4-1}
\usepackage{amsmath, amsthm, amssymb, mathtools}
\usepackage{enumitem}

\usepackage{ifpdf}
\ifpdf
\usepackage[pdftex]{graphicx}
\else
\usepackage[dvips]{graphicx}
\fi
\usepackage{tikz}
 	 \usetikzlibrary{arrows,backgrounds}
\usepackage[all]{xy}
\usepackage{tocvsec2}
\usepackage{comment}
\usepackage{bbm}
\usepackage{tabularx}
\usepackage{stmaryrd}

\input xy
\xyoption{all}

\usepackage[pdftex,plainpages=false,hypertexnames=false,colorlinks,pdfpagelabels]{hyperref}
\newcommand{\arxiv}[1]{\href{http://arxiv.org/abs/#1}{\tt arXiv:\nolinkurl{#1}}}
\newcommand{\arXiv}[1]{\href{http://arxiv.org/abs/#1}{\tt arXiv:\nolinkurl{#1}}}
\renewcommand{\doi}[1]{\href{http://dx.doi.org/#1}{{\tt DOI:#1}}}

\newcommand{\googlebooks}[1]{(preview at \href{http://books.google.com/books?id=#1}{google books})}

\usepackage{xcolor}
\definecolor{dark-red}{rgb}{0.7,0.25,0.25}
\definecolor{dark-blue}{rgb}{0.15,0.15,0.55}
\definecolor{medium-blue}{rgb}{0,0,.8}
\definecolor{violet}{RGB}{138,43,226}
\definecolor{DarkGreen}{RGB}{0,150,0}
\definecolor{rufous}{HTML}{a81c07}
\definecolor{Forestgreen}{rgb}{0.0, 0.5, 0.0}
\definecolor{darkaquamarine}{rgb}{.12, .47, .42}

\definecolor{cb-black}      {RGB}{  0,   0,   0}
\definecolor{cb-blue-green} {RGB}{  0,  073,  073}
\definecolor{cb-green-sea}  {RGB}{  0, 146, 146}
\definecolor{cb-rose}       {RGB}{255, 109, 182}
\definecolor{cb-salmon-pink}{RGB}{255, 182, 119}
\definecolor{cb-purple}     {RGB}{ 73,   0, 146}
\definecolor{cb-blue}       {RGB}{ 0, 109, 219}
\definecolor{cb-lilac}      {RGB}{182, 109, 255}
\definecolor{cb-blue-sky}   {RGB}{109, 182, 255}
\definecolor{cb-blue-light} {RGB}{182, 219, 255}
\definecolor{cb-burgundy}   {RGB}{146,   0,   0}
\definecolor{cb-brown}      {RGB}{146,  73,   0}
\definecolor{cb-clay}       {RGB}{219, 209,   0}
\definecolor{cb-green-lime} {RGB}{ 36, 255,  36}
\definecolor{cb-yellow}     {RGB}{255, 255, 109}

\hypersetup{
   linkcolor={purple},
   citecolor={medium-blue}, urlcolor={medium-blue}
}

\usepackage{fullpage}

\theoremstyle{plain}
\newtheorem{thm}{Theorem}[section]
\newtheorem*{thm*}{Theorem}

\newtheorem*{cor*}{Corollary}

\newtheorem*{conj*}{Conjecture}

\newtheorem*{quest*}{Question}
\newtheorem*{claim*}{Claim}

\theoremstyle{definition}
\newtheorem{defn}[thm]{Definition}

\newtheorem{sub-ex}[thm]{Sub-Example}
\newtheorem{rem}[thm]{Remark}

\newtheorem*{rem*}{Remark}

\DeclareMathOperator{\End}{End}

\DeclareMathOperator{\Hom}{Hom}

\DeclareMathOperator{\op}{op}
\DeclareMathOperator{\ONB}{ONB}

\DeclareMathOperator{\id}{id}
\DeclareMathOperator{\Id}{\Id}

\DeclareMathOperator{\im}{im}
\DeclareMathOperator{\Irr}{Irr}

\DeclareMathOperator{\Tr}{Tr}
\DeclareMathOperator{\tr}{tr}

\DeclareMathOperator{\Tube}{Tube}

\newcommand{\set}[2]{\left\{#1 \middle| #2\right\}}

\def\semicolon{;}
\def\applytolist#1{
    \expandafter\def\csname multi#1\endcsname##1{
        \def\multiack{##1}\ifx\multiack\semicolon
            \def\next{\relax}
        \else
            \csname #1\endcsname{##1}
            \def\next{\csname multi#1\endcsname}
        \fi
        \next}
    \csname multi#1\endcsname}

\def\calc#1{\expandafter\def\csname c#1\endcsname{{\mathcal #1}}}
\applytolist{calc}QWERTYUIOPLKJHGFDSAZXCVBNM;
\def\bbc#1{\expandafter\def\csname bb#1\endcsname{{\mathbb #1}}}
\applytolist{bbc}QWERTYUIOPLKJHGFDSAZXCVBNM;
\def\bfc#1{\expandafter\def\csname bf#1\endcsname{{\mathbf #1}}}
\applytolist{bfc}QWERTYUIOPLKJHGFDSAZXCVBNM;
\def\sfc#1{\expandafter\def\csname s#1\endcsname{{\sf #1}}}
\applytolist{sfc}QWERTYUIOPLKJHGFDSAZXCVBNM;
\def\fc#1{\expandafter\def\csname f#1\endcsname{{\mathfrak #1}}}
\applytolist{fc}QWERTYUIOPLKJHGFDSAZXCVBNM;
\def\rmc#1{\expandafter\def\csname rm#1\endcsname{{\mathrm #1}}}
\applytolist{rmc}QWERTYUIOPLKJHGFDSAZXCVBNM;

\newcommand{\Rep}{{\sf Rep}}

\newcommand{\Mod}{{\sf Mod}}

\newcommand{\Hilb}{{\sf Hilb}}

\newcommand{\noshow}[1]{}
\newcommand{\MR}[1]{}

\usetikzlibrary{shapes}
\usetikzlibrary{cd}
\usetikzlibrary{backgrounds}
\usetikzlibrary{intersections}
\usetikzlibrary{decorations,decorations.pathreplacing,decorations.markings,decorations.pathmorphing}
\usetikzlibrary{fit,calc,through}
\usetikzlibrary{external}
\usetikzlibrary{arrows}
\tikzset{vertex/.style = {shape=circle,draw,fill=black,inner sep=0pt,minimum size=5pt}}
\tikzset{edge/.style = {->,> = latex', bend right}}
\tikzset{
	super thick/.style={line width=3pt}
}
\tikzstyle{knot}=[preaction={super thick, white, draw}]

\tikzset{
    quadruple/.style args={[#1] in [#2] in [#3] in [#4]}{
        #1,preaction={preaction={preaction={draw,#4},draw,#3}, draw,#2}
    }
}
\tikzstyle{shaded}=[fill=red!10!blue!20!gray!30!white]
\tikzstyle{unshaded}=[fill=white]
\tikzstyle{empty box}=[circle, draw, thick, fill=white, opaque, inner sep=2mm]
\tikzstyle{annular}=[scale=.7, inner sep=1mm, baseline]
\tikzstyle{rectangular}=[scale=.75, inner sep=1mm, baseline=-.1cm]
\tikzstyle{mid>}=[decoration={markings, mark=at position 0.53 with {\arrow{>}}}, postaction={decorate}]
\tikzstyle{mid<}=[decoration={markings, mark=at position 0.5 with {\arrow{<}}}, postaction={decorate}]
\tikzstyle{over}=[double, draw=white, super thick, double=]
\tikzstyle{box} = [rectangle,draw,rounded corners=5pt,very thick]

\newcommand{\roundNbox}[6]{
	\draw[rounded corners=5pt, very thick, #1] ($#2+(-#3,-#3)+(-#4,0)$) rectangle ($#2+(#3,#3)+(#5,0)$);
	\coordinate (ZZa) at ($#2+(-#4,0)$);
	\coordinate (ZZb) at ($#2+(#5,0)$);
	\node at ($1/2*(ZZa)+1/2*(ZZb)$) {#6};
}

\newcommand{\halfDottedEllipse}[4][]{
	\draw[#1, thick] #2 arc(-180:0:{#3} and {#4});
	\draw[#1, thick, dotted] ($ #2 + 2*(#3,0)$) arc(0:180:{#3} and {#4});
}
\newcommand{\halfDottedEllipseTwo}[4][]{
	\draw[#1, thick, dotted] #2 arc(-180:0:{#3} and {#4});
	\draw[#1, thick] ($ #2 + 2*(#3,0)$) arc(0:180:{#3} and {#4});
}

\newcommand{\pairOfPantsTwo}[1]{
	\draw[thick] #1 .. controls ++(90:.8cm) and ++(270:.8cm) .. ($ #1 + (.7,1.5) $);
	\draw[thick] ($ #1 + (2,0) $) .. controls ++(90:.8cm) and ++(270:.8cm) .. ($ #1 + (2,0) + (-.7,1.5) $);
	\draw[thick] ($ #1 + (.6,0) $).. controls ++(90:.8cm) and ++(90:.8cm) .. ($ #1 + (1.4,0) $); 
	\halfDottedEllipseTwo{($ #1 + (.7,1.5) $)}{.3}{.1}
	\halfDottedEllipseTwo{#1}{.3}{.1}
	\halfDottedEllipseTwo{($ #1 + (1.4,0) $)}{.3}{.1}
}
\newcommand{\topPairOfPants}[1]{
	\draw[thick] #1 .. controls ++(90:.8cm) and ++(270:.8cm) .. ($ #1 + (.7,1.5) $);
	\draw[thick] ($ #1 + (2,0) $) .. controls ++(90:.8cm) and ++(270:.8cm) .. ($ #1 + (2,0) + (-.7,1.5) $);
	\draw[thick] ($ #1 + (.6,0) $).. controls ++(90:.8cm) and ++(90:.8cm) .. ($ #1 + (1.4,0) $); 
	\draw[thick] ($ #1 + (1,1.5) $) ellipse (.3cm and .1cm);
	\halfDottedEllipse{#1}{.3}{.1}
	\halfDottedEllipse{($ #1 + (1.4,0) $)}{.3}{.1}
}

\newcommand{\topPairOfPantsTwo}[1]{
	\draw[thick] #1 .. controls ++(90:.8cm) and ++(270:.8cm) .. ($ #1 + (.7,1.5) $);
	\draw[thick] ($ #1 + (2,0) $) .. controls ++(90:.8cm) and ++(270:.8cm) .. ($ #1 + (2,0) + (-.7,1.5) $);
	\draw[thick] ($ #1 + (.6,0) $).. controls ++(90:.8cm) and ++(90:.8cm) .. ($ #1 + (1.4,0) $); 
	\halfDottedEllipseTwo{($ #1 + (.7,1.5) $)}{.3}{.1}
\draw[thick] ($ #1 + (.3,0)$) ellipse (.3 and .1);
\draw[thick] ($ #1 + (1.7,0)$) ellipse (.3 and .1);
}

\newcommand{\emptyCylinder}[3]{
	\draw[thick] #1 -- ($ #1 + (0,#3) $);
	\draw[thick] ($ #1 + 2*(#2,0) $) -- ($ #1 + 2*(#2,0) + (0,#3) $);	
}

\newcommand{\topCylinderTwo}[2]{
	\draw[thick] #1 -- ($ #1 + (0,1) $);
	\draw[thick] ($ #1 + (.6,0) $) -- ($ #1 + (.6,1) $);
	\halfDottedEllipseTwo{($ #1 + (0,1) $)}{.3}{.1}
}

\newcommand{\bottomCylinderTwo}[3]{
	\draw[thick] #1 -- ($ #1 + (0,#3) $);
	\draw[thick] ($ #1 + 2*(#2,0) $) -- ($ #1 + 2*(#2,0) + (0,#3) $);
	\halfDottedEllipseTwo{#1}{#2}{{1/3*#2}}	
}

\newcommand{\colorHalfDottedEllipseTwo}[4]{
	\draw[thick, dotted, #4] #1 arc(-180:0:{#2} and {#3});
	\draw[thick, #4] ($ #1 + 2*(#2,0)$) arc(0:180:{#2} and {#3});
}

\newcommand{\tikzmath}[2][]{\vcenter{\hbox{\begin{tikzpicture}[#1]#2
\end{tikzpicture}}}
}

\newcommand{\Xstring}{red}
\newcommand{\Cstring}{blue}
\newcommand{\CstringOne}{cyan}
\newcommand{\Zstring}{orange}

\begin{document}


\title{Levin-Wen is a gauge theory: entanglement from topology}

\author{Kyle Kawagoe${}^{1,3}$}
\author{Corey Jones${}^2$}
\author{Sean Sanford${}^3$}
\author{David Green${}^3$}
\author{David Penneys${}^3$}
\affiliation{${}^1$Department of Physics, The Ohio State University, Columbus, OH 43212, USA}
\affiliation{${}^2$Department of Mathematics, North Carolina State University, Raleigh, North Carolina 27695, USA}
\affiliation{${}^3$Department of Mathematics, The Ohio State University, Columbus, OH 43210, USA}

\begin{abstract}
We show that the Levin-Wen model of a unitary fusion category $\mathcal{C}$ is a gauge theory with gauge symmetry given by the tube algebra $\operatorname{Tube}(\mathcal{C})$.
In particular, we define a model corresponding to a $\operatorname{Tube}(\mathcal{C})$ symmetry protected topological phase, and we provide a gauging procedure which results in the corresponding Levin-Wen model. 
In the case $\mathcal{C}=\mathsf{Hilb}(G,\omega)$, we show how our procedure reduces to the twisted gauging of a trivial $G$-SPT to produce the Twisted Quantum Double. 
We further provide an example which is outside the bounds of the current literature, the trivial Fibonacci SPT, whose gauge theory results in the doubled Fibonacci string-net. 
Our formalism has a natural topological interpretation with string diagrams living on a punctured sphere. 
We provide diagrams to supplement our mathematical proofs and to give the reader an intuitive understanding of the subject matter.
\end{abstract}

\maketitle
\section{Introduction}

In the original string-net paper \cite{PhysRevB.71.045110}, Levin and Wen argue that many lattice gauge theories in 2+1D may be realized by string-nets.
String-net models, in addition to constructing gauge theories, generate a massive class of 2+1D topological phases of matter corresponding to the Drinfeld centers of unitary fusion categories (UFCs).
In this article, we show that all Levin-Wen models are themselves gauge theories. 
We introduce a trivial $\Tube(\cC)$ symmetry protected topological (SPT) phase for any UFC $\cC$ and show how to gauge it to produce the Levin-Wen model for $\cC$ whose anyonic excitations are described by the Drinfeld center $Z(\cC)$.

In both field theories and lattice models, many topologically ordered phases of matter are gauge theories which may be obtained by gauging a theory with a global symmetry
\cite{RevModPhys.51.659}.
For example, the Twisted Quantum Double of a finite group $G$ may be constructed by gauging a non-trivial $G$-SPT with global symmetry group $G$
\cite{PhysRevB.86.115109,PhysRevB.87.125114,PhysRevB.108.115144}.
Alternatively, the Twisted Quantum Double also arises from a trivial $G$-SPT via a twisted gauging procedure, similar to a twisted Kramers-Wannier tranformation \cite{2112.01519}.
The eigenvalue equations of the plaquette operators of a lattice gauge theory are naturally interpreted as Gauss laws measuring local charge
\cite{RevModPhys.51.659,MR1951039}\footnote{We are considering a theory defined on the dual lattice of the lattice considered by \cite{MR1951039}.}.
Likewise, although non-abelian symmetries require more finesse, the vertex operators measure magnetic flux and assign to the Hamiltonian the appropriate energy cost. 
It is therefore reasonable to expect that the vertex and plaquette operators of the Levin-Wen model have similar interpretation in terms of a gauge symmetry. 
One of the main results of this paper is to identify this gauge symmetry.

It is known that the tube algebra $\Tube(\cC)$ acts locally on the Levin-Wen Hilbert space and is able to distinguish between its anyonic excitations in $Z(\cC)$ and leaves the ground state invariant
\cite{MR3614057,MR4642306}. 
Mathematically, this is summarized in the statement $\Mod(\Tube(\cC))\cong Z(\cC)$ as unitary categories \cite{MR1782145,MR1832764,MR1966525}. 
This suggests that $\Tube(\cC)$ might be a gauge symmetry. 
In order to establish that $\Tube(\cC)$ is a gauge symmetry, we construct a $\Tube(\cC)$-symmetric theory which becomes the $Z(\cC)$ Levin-Wen model when gauged.

Unlike $\mathbb{C}[G]$ for a finite group $G$, the algebra $\Tube(\cC)$ is not a (weak) Hopf algebra and does not admit a comultiplication inducing a tensor product on its representations; indeed, $Z(\cC)$ does not typically admit a fiber functor to $\Hilb$.
Physically, this means that there are no lattice models where $\Tube(\cC)$ is realized as a global on-site symmetry on the local Hilbert space.  
Instead, we perform the tensor product of $\Tube(\cC)$-modules under the equivalence $\Mod(\Tube(\cC))\cong Z(\cC)$, which corresponds to a Day convolution product \cite{MR0272852} over sites.  
This convolution product has a global $\Tube(\cC)$-action. 

At this point, one may object that this Hilbert space is not a tensor product, and thus cannot host a local Hamiltonian, let alone a short range entangled ground state characteristic of SPT models.
However, we show in \S\ref{sec:LocalOperators} that the operators in our model carry a local structure given by a local net of algebras, and the observable quantities of the theory are consistent with the locality imposed by the lattice.

In the case of gauging a finite group $G$, one starts with a short range entangled theory. However, in the initial Hilbert space of our gauging procedure, the usual definition of short range entanglement does not make sense.  Instead, we start with a symmetric theory with a unique gapped ground state, even on a finite lattice.  We take this as a signature of a lack of topological order.

We now give a heuristic description of our $\Tube(\cC)$ symmetric theory, the $\Tube(\cC)$ symmetry action, and the gauging procedure which produces the Levin-Wen model. 
We build our lattice on a large disk, where each lattice site is a small punctured disk inside.
The Hilbert space on each lattice site is given by
$$
\tikzmath{
\draw[thick, dashed] (0,0) circle (1cm);
\node at (0,0) {$\times$};
\draw[thick, blue, mid>] (0,-.5) arc (270:-90:.5cm);
\draw[thick, blue, mid<] (0,-.5) --node[right]{$\scriptstyle b$} (0,-1);
\filldraw[blue] (0,-.5) node[above]{$\scriptstyle \gamma$} circle (.05cm);
\node[blue] at (.65,0) {$\scriptstyle \overline{d}$};
\node[blue] at (-.65,0) {$\scriptstyle d$};
}
\qquad
\bigoplus_{b,d\in\Irr(\cC)}
\begin{aligned}
\{
\gamma \in \cC(b\to &d\otimes \overline{d})
|
\\
b,d&\in\Irr(\cC)
\}.
\end{aligned}
$$
One should consider this simple object as labeling the \emph{domain state} on the disk. 
A disk with label $d\in \Irr(\cC)$ may have a twisted \emph{boundary condition} labeled by a simple object $b\in \Irr(\cC)$ and a basis vector in $\cC(b\rightarrow d\otimes \overline{d})$. 
The on-site Hilbert space carries a local right $\Tube(\cC)$-action which can change the boundary conditions.
This action will eventually be used to define the Hamiltonian on each site.
When $b=1_\cC$, we say the boundary is \emph{untwisted} and when $b\neq 1_\cC$, we say the boundary is \emph{twisted} by $b$; in the latter case, there is a $b$-string emanating from the small disk into the interstitial space between the small disks on the lattice. 
The remaining degrees of freedom are related to how these strings fuse with one another. 
We allow a single string to terminate on the boundary of the large disk of the lattice, which affords the global $\Tube(\cC)$-action. 

The ground state is the linear combination
$$
\frac{1}{D_\cC}
\sum_{d\in\Irr(\cC)}
d_d
\cdot
\tikzmath{
\draw[thick, dashed] (0,0) circle (.8cm);
\node at (0,0) {$\times$};
\draw[thick, blue, mid>] (0,-.5) arc (270:-90:.5cm);
\node[blue] at (.65,0) {$\scriptstyle \overline{d}$};
\node[blue] at (-.65,0) {$\scriptstyle d$};
}
$$
on each site,
where $D_\cC$ is the global dimension of $\cC$.
This weighted sum over all untwisted domain states screens the local puncture so that the global ground state is just a large disk with no punctures.
The global action of the tube algebra on the ground state is trivial.  

Just as in the usual gauging procedure which generates the (Twisted) Quantum Double, we may imagine the small punctured disks on sites growing so that the domains are touching, which creates domain walls.  
Of course, the $b$-strings which get stuck between these domains will influence the domain wall states. 
These domain walls are now \emph{string-like} and their locations and fusions will create some degrees of freedom where domain walls meet.
In the body of the article, we will see the resulting space is the intermediate Levin-Wen Hilbert space (the ground state space of the edge/vertex terms), and the Hamiltonian on the original model becomes the charge part of the Levin-Wen Hamiltonian (the sum of the plaquette terms).
The vertex part of the Levin-Wen Hamiltonian is the gauge constraint.

In \S\ref{sec:LWStringNet}, we give an overview of the Levin-Wen string net model from a unitary tensor category perspective rather than the usual 6j symbol viewpoint which was given in \cite{2305.14068} based on \cite{MR3204497}.
In \S\ref{sec:CatTrace}, we discuss two main tools which we use in our gauging procedure: the categorified trace $\cC\to Z(\cC)$ \cite{MR3578212}, a.k.a.~the adjoint to the forgetful functor $Z(\cC)\to \cC$ \cite{MR1966525}, and Ocneanu's tube algebra \cite{MR1782145,MR1832764,MR1966525}.
The main construction of the paper occurs in \S\ref{sec:UngaugingLevinWen}.
The explicit gauging map is constructed in \S\ref{sec:TheGaugingMap}, and we discuss the (quasi-)local algebras of observables in \S\ref{sec:LocalOperators} and \S\ref{sec:QuasiLocalOperators}.
We give a modified model in \S\ref{sec:ModifiedModel} which can see all excitations including fluxes and dyons, and not just charges; this modified model also has the benefit that $\Tube(\cC)$ now acts locally on-site as a local gauge symmetry action.
We compute explicit group examples in \S\ref{sec:GroupExample}, showing our construction generalizes the usual story of gauging a (twisted) $G$-SPT, and in \S\ref{sec:FibExample}, we gauge the trivial Fibonacci SPT to obtain the Fibonacci Levin-Wen theory.


\section{Levin-Wen string net model}
\label{sec:LWStringNet}

We begin by rapidly recalling the unitary tensor category version of the Levin-Wen model from \cite{2305.14068} based on \cite{MR3204497}, as opposed to the usual skeletal 6j symbol version from \cite{PhysRevB.71.045110,PhysRevB.103.195155}.
Let $\cC$ be a UFC.
We define the distinguished object $X\coloneqq\bigoplus_{c\in\Irr(\cC)}  c$, which we represent by a red strand, and we denote the space $\cC(X \otimes X\to X \otimes X)$
by a single red 4-valent vertex, which should be viewed as going from left-to-right and bottom-to-top.
\begin{align*}
\tikzmath{
\draw[dotted, rounded corners=5pt] (0,0) rectangle (.6,.6);
\draw[thick, red] (.3,0) -- (.3,.6);
}
&=
X
=
\bigoplus_{c\in\Irr(\cC)}  c
\\
\tikzmath{
\draw[blue!30] (.1,.5) -- (.5,.1);
\draw[blue!30,->] (.1,.4) -- (.25,.55);
\draw[blue!30,->] (.4,.1) -- (.55,.25);
\draw[dotted, rounded corners=5pt] (0,0) rectangle (.6,.6);
\fill[red] (.3,.3) circle (.05cm);
\draw[thick, red] (.3,0) -- (.3,.6);
\draw[thick, red] (0,.3) -- (.6,.3);
}
&=
\cC(X\otimes  X\to X\otimes  X)
\\&=
\bigoplus_{\substack{a,b,c,d\\\in\Irr(\cC)}}\cC(a \otimes b\to c \otimes d).
\end{align*}
This 4-valent space carries the \emph{skein module inner product}.
When the simple labels do not match, the inner product is zero, and when they do, the inner product is given by
$$
\left\langle
\tikzmath{
\draw (-.5,0) node[left]{$\scriptstyle a$} -- (.5,0) node[right]{$\scriptstyle d$};
\draw (0,-.5) node[below]{$\scriptstyle b$} -- (0,.5) node[above]{$\scriptstyle c$};
\roundNbox{fill=white}{(0,0)}{.2}{0}{0}{$\scriptstyle f$}
}
\Bigg|
\tikzmath{
\draw (-.5,0) node[left]{$\scriptstyle a$} -- (.5,0) node[right]{$\scriptstyle d$};
\draw (0,-.5) node[below]{$\scriptstyle b$} -- (0,.5) node[above]{$\scriptstyle c$};
\roundNbox{fill=white}{(0,0)}{.2}{0}{0}{$\scriptstyle g$}
}
\right\rangle
\coloneqq
\frac{\tr_\cC(f^\dag g)}{
\sqrt{d_ad_bd_cd_d}
}
$$
where $\tr_\cC$ is the categorical unitary spherical trace.

The local Hilbert space is $\cH_v=\cC(X\otimes X\to X\otimes X)$ with the above inner product, and the microscopic Hilbert space is $\cH=\bigotimes_{v} \cH_v$.
$$
\tikzmath{
\foreach \x in{0,.7,1.4,2.1}{
\foreach \y in{0,.7,1.4}{
\fill[red] (\x,\y) circle (.05cm);
\draw[thick, red] ($(\x,\y)-(.3,0)$) -- ($(\x,\y)+(.3,0)$);
\draw[thick, red] ($(\x,\y)-(0,.3)$) -- ($(\x,\y)+(0,.3)$);
}}
}
$$
The commuting projector local Hamiltonian is made of edge terms $A_\ell$, which force the simples labelling edges to match, and plaquette terms $B_p$ defined on $P_A(\bigotimes_{v} \cH_v)$ where $P_A=\prod_{\ell} A_\ell$.
The operator $B_p$ glues in the regular element and resolves the morphism back into the tensor product.
$$
B_p\tikzmath{
\foreach \x in{0,.7}{
\foreach \y in{0,.7}{
\fill[red] (\x,\y) circle (.05cm);
\draw[thick, red] ($(\x,\y)-(.35,0)$) -- ($(\x,\y)+(.35,0)$);
\draw[thick, red] ($(\x,\y)-(0,.35)$) -- ($(\x,\y)+(0,.35)$);
}}
} = 
\frac{1}{D_\cC}
\sum_{c \in \Irr(\cC)} 
d_c\,
\tikzmath{
\foreach \x in{0,.7}{
\foreach \y in{0,.7}{
\fill[red] (\x,\y) circle (.05cm);
\draw[thick, red] ($(\x,\y)-(.35,0)$) -- ($(\x,\y)+(.35,0)$);
\draw[thick, red] ($(\x,\y)-(0,.35)$) -- ($(\x,\y)+(0,.35)$);
}}
\draw[thick, blue, rounded corners=5pt] (.1,.1) rectangle (.6,.6);
\node[blue] at (.2,.35) {$\scriptstyle c$};
}
$$
There are elegant formulas for $\im(P_BP_A)$ and $\im(P_A)$ in terms of skein modules.
The first was announced in \cite{MR3204497}, and proofs of both can be found in \cite[\S2.2]{2305.14068}.
\begin{itemize}
\item 
We can identify the ground state space as a skein module under a unitary isomorphism
$$
\im(P_BP_A)
\cong
\cC(X^{\otimes \partial \Lambda}\to 1_\cC).
$$
Here, $X^{\otimes \partial \Lambda}$ means take the tensor product of $X$ over the boundary sites of $\Lambda$.
\item
On a contractible rectanglar patch $\Lambda$ of lattice, we have a unitary isomorphism
\begin{equation}
\label{eq:IntermediateSpace}
\im(P_A)
\cong
\cC(X^{\otimes \partial \Lambda} \to F(\Tr(1_\cC))^{\otimes p\in\Lambda}).
\end{equation}
Here, $X^{\otimes \partial \Lambda}$ is as above, $\otimes p\in\Lambda$ means the tensor product over the plaquettes in $\Lambda$ (this is made more precise in \S\ref{sec:SpineCoralSkeinModule}), $F:Z(\cC)\to \cC$ is the forgetful functor, and $\Tr:\cC\to Z(\cC)$ is its unitary adjoint \cite[Def.~2.2]{2301.11114}.
We provide a more detailed treatment of the unitary adjoint $\Tr_\cC$ in the next section.

This space again carries the skein module inner product, where we must expand
$$
F(\Tr(1_\cC)) = \bigoplus_{c\in\Irr(\cC)} c\otimes \overline{c}.
$$
Note that $\Tr(1_\cC)\in Z(\cC)$ is the canonical Lagrangian algebra $A\in Z(\cC)$ corresponding to $\cC$ as a gapped boundary to the vacuum.
In Section \ref{sec:UngaugingHilbertSpace} below, we will describe how each plaquette operator $B_p$ acts on the the right hand side of the unitary isomorphism \eqref{eq:IntermediateSpace}.
\end{itemize}

\section{Categorified trace and the tube algebra}
\label{sec:CatTrace}

We now review two important ingredients for our construction: the categorified trace and the tube algebra.

\subsection{Categorified trace}

In \cite{MR3578212}, it was shown that the right adjoint of a braided pivotal functor
carries the structure of a \emph{categorified trace} and can be represented by a graphical calculus of strings on tubes.
In particular, the right adjoint to the forgetful functor $F:Z(\cC)\to \cC$ carries this structure, which we denote by $\Tr:\cC\to Z(\cC)$, and which is represented graphically (reading bottom to top) by
$$
\Tr(f:a\to b)
=
\tikzmath{
\draw[thick] (0,0) -- (0,1.3);
\draw[thick] (.9,0) -- (.9,1.3);
\filldraw[thick, fill=white] (.45,1.3) ellipse (.45 and .15);
\draw[thick, blue] (.45,-.15) node[below]{$\scriptstyle a$} -- (.45,.3);
\draw[thick, blue] (.45,1.15) node[above,yshift=-.05cm]{$\scriptstyle b$} -- (.45,.8);
\halfDottedEllipse{(0,0)}{.45}{.15}
\roundNbox{fill=white}{(.45,.55)}{.25}{0}{0}{$f$}
}\,.
$$
We will typically draw strings from $\cC$ in blue to distinguish them from strings in $Z(\cC)$, which we draw in orange with texture indicative of an `excitation.'

We denote the unit of the adjunction $\eta_z:z\to \Tr(F(z))$ by a cup where a string from $Z(\cC)$ can pass on to the tube, and we write $i\coloneqq \eta_1$.
$$
\eta_z 
=
\tikzmath{
\draw[thick] (0,.5) -- (0,0) arc (-180:0:.45cm) -- (.9,.5);
\filldraw[thick, fill=white] (.45,.5) ellipse (.45 and .15);
\draw[thick, knot, \Zstring, decorate, decoration={snake, segment length=1mm, amplitude=.2mm}] (.15,-.8)  node[right, yshift=.1cm]{$\scriptstyle z$} to[out=90,in=270] (.45,-.3) -- (.45,.35);
}
\qquad\qquad
i=
\tikzmath{
\draw[thick] (0,.5) -- (0,0) arc (-180:0:.45cm) -- (.9,.5);
\filldraw[thick, fill=white] (.45,.5) ellipse (.45 and .15);
}
$$
While we will not explicitly need the counit of the adjunction, it is worth mentioning that
$$
F(\Tr(a)) = \bigoplus_{c\in\Irr(\cC)} c\otimes a\otimes \overline{c}.
$$

As $F$ is monoidal, the right adjoint is lax monoidal, endowing it with a canonical `multiplication' which is represented by a pair of pants.
$$
\mu_{a,b}
=
\tikzmath{
	\coordinate (b1) at (0,.3);
	\coordinate (b2) at (1.4,.3);
	\coordinate (c1) at (1,1.7);
	\topPairOfPants{(b1)}{}
	\draw[thick, blue] ($ (b1) + (.3,-.1) $) node[below]{$\scriptstyle a$} .. controls ++(90:.5cm) and ++(270:.4cm) .. ($ (c1) - (.1,0) $);
	\draw[thick, blue] ($ (b2) + (.3,-.1) $) node[below]{$\scriptstyle b$} .. controls ++(90:.5cm) and ++(270:.4cm) .. ($ (c1) + (.1,0) $);
}
$$
By the results of \cite{MR4528312,2301.11114}, string diagrams on tubes which branch and braid are invariant up to isotopy fixing the boundary circles, and there are several relations involving braiding and twists in $Z(\cC)$; we omit these here as they are not necessary for this article.

Now since $\cC$ and $Z(\cC)$ are endowed with their canonical unitary spherical structures, the adjunction isomorphism
$$
\cC(F(z)\to c) \cong Z(\cC)(z\to \Tr(c))
$$
is unitary; this is the definition of a \emph{unitary adjunction}.
By \cite[\S2.3-4]{2301.11114}, $\Tr:\cC\to Z(\cC)$ is a lax monoidal dagger functor which carries a canonical unitary involutive structure compatible with the involutions in $\cC$ and $Z(\cC)$.
Moreover, $\Tr$ is automatically a 2-sided adjoint of $F$, as we may take dagger uniformly for all the above maps.
We will do this below as we wish to discuss \emph{right modules} of Ocneanu's tube algebra.

We remark that by 
\cite[Thm.~4.4 and Rem~4.6]{MR3663592} (see also
\cite[Rem.~2.13]{2305.14068}), all morphisms in $Z(\cC)(\Tr(a)\to \Tr(b))$ can be realized as string diagrams on tubes as
\begin{align*}
Z(\cC)(\Tr(a)\to \Tr(b))
&\cong
\cC(F(\Tr(a))\to b)
\\&=
\bigoplus_{c\in\Irr(\cC)}\cC(c\otimes a\otimes \overline{c}\to b)
\\&\cong
\bigoplus_{c\in\Irr(\cC)}\cC(c\otimes a\to b\otimes c),
\end{align*}
and we can identify $Z(\cC)(\Tr(a)\to \Tr(b))$
with
$$
\tikzmath{
\draw[thick] (0,0) -- (0,1);
\draw[thick] (1,0) -- (1,1);
\draw[thick, blue] (.5,.3) --node[right]{$\scriptstyle a$} (.5,-.2);
\draw[thick, blue] (.5,.3) --node[right]{$\scriptstyle b$} (.5,.8);
\halfDottedEllipse{(0,0)}{.5}{.2}
\halfDottedEllipse[thick, red]{(0,.5)}{.5}{.2}
\filldraw[thick, fill=white] (.5,1) ellipse (.5 and .2);
\fill[red] (.5,.3) circle (.05cm);
}
\coloneqq
\set{
\tikzmath{
\draw[thick] (0,0) -- (0,1);
\draw[thick] (1,0) -- (1,1);
\draw[thick, blue] (.5,.3) --node[right]{$\scriptstyle a$} (.5,-.2);
\draw[thick, blue] (.5,.3) --node[right, yshift=.05cm]{$\scriptstyle b$} (.5,.8);
\halfDottedEllipse{(0,0)}{.5}{.2}
\halfDottedEllipse[thick, red]{(0,.5)}{.5}{.2}
\filldraw[thick, fill=white] (.5,1) ellipse (.5 and .2);
\fill[red] (.5,.3) circle (.05cm);
\roundNbox{fill=white}{(.5,.3)}{.2}{0}{0}{$\scriptstyle f$}
}
~}{~
f\in \cC(X\otimes a\to b\otimes X)
}.
$$

\subsection{Tube algebra}
\label{sec:TubeAlgebra}

Ocneanu's tube algebra
\begin{align*}
\Tube(\cC)
&=
\tikzmath{
\draw[thick] (0,0) -- (0,1);
\draw[thick] (1,0) -- (1,1);
\draw[thick, red] (.5,-.2) -- (.5,.8);
\halfDottedEllipse{(0,0)}{.5}{.2}
\halfDottedEllipse[thick, red]{(0,.5)}{.5}{.2}
\filldraw[thick, fill=white] (.5,1) ellipse (.5 and .2);
\fill[red] (.5,.3) circle (.05cm);
}
=
\End_{Z(\cC)}(\Tr(X))
\\&\cong
\bigoplus_{a,b,c\in\Irr(\cC)}\cC(c\otimes a\to b\otimes c)
\end{align*}
is a 
finite dimensional $\rmC^*$-algebra which is highly useful to compute $Z(\cC)$ \cite{MR1782145,MR1832764,MR1966525}.
The algebra structure is given by stacking tubes and resolving them using the graphical calculus via the \emph{fusion relation} 
\begin{equation}
\label{eq:FusionRelation}
\id_{a\otimes b}
=
\tikzmath{
\draw (-.2,.-.7) node[below]{$\scriptstyle a$} -- (-.2,.7);
\draw (.2,.-.7) node[below]{$\scriptstyle b$} -- (.2,.7);
}
=
\sum_{c\in \Irr(\cC)}
\sqrt{\frac{d_c}{d_ad_b}}
\sum_{\phi \in \cB_{c}^{ab}}
\tikzmath{
\draw (-.2,-1) node[left, yshift=.2cm]{$\scriptstyle a$} -- (-.2,-.6);
\draw (.2,-1) node[right, yshift=.2cm]{$\scriptstyle b$} -- (.2,-.6);
\draw (-.2,1) node[left, yshift=-.2cm]{$\scriptstyle a$} -- (-.2,.6);
\draw (.2,1) node[right, yshift=-.2cm]{$\scriptstyle b$} -- (.2,.6);
\draw (0,-.5) -- node[right]{$\scriptstyle c$} (0,.5);
\roundNbox{fill=white}{(0,.4)}{.2}{.2}{.2}{$\scriptstyle\phi$}
\roundNbox{fill=white}{(0,-.4)}{.2}{.2}{.2}{$\scriptstyle\phi^\dag$}
}
\end{equation}
where $\cB_c^{ab}$ is an ONB for $\cC(c\to a\otimes b)$ in the skein module inner product.
The $*$-structure is given by taking $\dag$ in $Z(\cC)$, which is compatible with $\dag$ in $\cC$ as $\Tr$ is a dagger functor.

\begin{defn}
\label{defn:TensorProduct}
A \emph{right module} for the tube algebra $\Tube(\cC)$ is a finite dimensional Hilbert space $\cK$ endowed with a right $\Tube(\cC)$-action, i.e.., a unital $*$-homomorphism $\pi_\cK: \Tube(\cC)^{\op}\to B(\cK)$.

By \cite[\S5-6]{MR1966525}, there is an equivalence of unitary categories $\Mod(\Tube(\cC))\cong Z(\cC)$, where on the right hand side, we take the underlying category of the UMTC $Z(\cC)$.
Starting with a right module $\cK$ for $\Tube(\cC)$, the projections $p_c:= \Tr(\id_c)$ for $c\in\Irr(\cC)$ endow the Hilbert space $\cK$ with an $\Irr(\cC)$-grading: $\cK_c:= p_c\cK$.
We get an object in $\cC$ by the formula
$$
K = \bigoplus_{c\in \Irr(\cC)} \cK_c \otimes c
$$
which has a canonical half-braiding and thus defines an object of $Z(\cC)$.
Conversely, starting with $K\in Z(\cC)$,
we define an $\Irr(\cC)$-graded Hilbert space $\cK$ by
$$
\cK_c := Z(\cC)(\Tr(c)\to K)\cong \cC(c\to F(K)),
$$ 
and we get a right $\Tube(\cC)$-action by precomposition with morphisms $\Tr(f)\in \Hom(\Tr(a)\to \Tr(b))$ 
for $f\in \cC(a\to b)$.
These two constructions are mutually inverse, and we refer the reader to \cite{MR1966525,MR4642306,2305.14068} for more details.

Passing back and forth between these perspectives allows us to view a right $\Tube(\cC)$-module $\cK$ as a functor $\cC^{\op}\to \Hilb$,
which affords a description of the tensor product of $\Tube(\cC)$-representations as the \emph{Day convolution product} \cite{MR0272852} (see also \cite{MR3254423,MR3687214}).
That is, given $\cH,\cK\in\Mod(\Tube(\cC))$ and $c\in\Irr(\cC)$, we define the $c$-graded component as
$$
(\cH\boxtimes \cK)_c = \bigoplus_{a,b\in \Irr(\cC)}  \cC(c\to a\otimes b)\otimes \cH_a\otimes \cK_b.
$$
Using the diagrammatic calculus of strings on tubes, one can view $(\cH\boxtimes \cK)_c$ as the linear span of
$$
\set{
\tikzmath[yscale=-1]{
	\coordinate (a1) at (0,0);
	\coordinate (a2) at (1.4,0);
	\coordinate (b1) at (0,.3);
	\coordinate (b2) at (1.4,.3);
	\coordinate (c1) at (1,1.2);
	\emptyCylinder{(a1)}{.3}{.3}
	\draw[thick] (a1) arc (-180:0:.3cm);
	\node at ($ (a1) + (.3,-.05) $) {$\eta$};
	\node at ($ (a1) + (.2,.55) $) {\scriptsize{$a$}};
 \draw[thick, \Zstring, decorate, decoration={snake, segment length=1mm, amplitude=.2mm}] ($ (a1) + (.3,-.3)$) --node[left]{$\scriptstyle H$} ($ (a1) + (.3,-.7)$);
	\emptyCylinder{(a2)}{.3}{.3}
	\draw[thick] (a2) arc (-180:0:.3cm);
	\node at ($ (a2) + (.3,-.05) $) {$\xi$};
	\node at ($ (a2) + (.4,.55) $) {\scriptsize{$b$}};
 \draw[thick, \Zstring, decorate, decoration={snake, segment length=1mm, amplitude=.2mm}] ($ (a2) + (.3,-.3)$) --node[left]{$\scriptstyle K$} ($ (a2) + (.3,-.7)$);
	\pairOfPantsTwo{(b1)}{}
	\draw[thick, blue] (c1) -- (1,1.9);
	\draw[thick, blue] ($ (b1) + (.3,.1) $) .. controls ++(90:.5cm) and ++(225:.4cm) .. (c1);
	\draw[thick, blue] ($ (b2) + (.3,.1) $) .. controls ++(90:.5cm) and ++(-45:.4cm) .. (c1);
	\filldraw (c1) circle (.05cm);
	\node at ($ (c1) + (-.2,.05) $) {\scriptsize{$\psi$}};
	\node at ($ (c1) + (.15,.3) $) {\scriptsize{$c$}};
}
\,
~}
{~
\begin{aligned}
\eta&\in \cH_a=Z(\cC)(\Tr(a)\to H)
\\
\xi&\in \cK_b=Z(\cC)(\Tr(b)\to K)
\\
\psi&\in \cC(c\to a\otimes b)
\end{aligned}
}
\,.
$$

For $f\in Z(\cC)(\Tr(d)\to \Tr(c))\subset p_c\Tube(\cC)p_d$, we define the action $(\cH\boxtimes \cK)_c \to (\cH\boxtimes \cK)_d$ as follows.
\[
\tikzmath[yscale=-1]{
	\coordinate (a1) at (0,0);
	\coordinate (a2) at (1.4,0);
	\coordinate (b1) at (0,.3);
	\coordinate (b2) at (1.4,.3);
	\coordinate (c1) at (1,1.2);
	\coordinate (d1) at (.7,1.8);
	\coordinate (e1) at (.7,2.25);
	\emptyCylinder{(a1)}{.3}{.3}
	\draw[thick] (a1) arc (-180:0:.3cm);
	\node at ($ (a1) + (.3,-.05) $) {$\eta$};
	\node at ($ (a1) + (.2,.55) $) {\scriptsize{$a$}};
 \draw[thick, \Zstring, decorate, decoration={snake, segment length=1mm, amplitude=.2mm}] ($ (a1) + (.3,-.3)$) --node[left]{$\scriptstyle H$} ($ (a1) + (.3,-.7)$);
	\emptyCylinder{(a2)}{.3}{.3}
	\draw[thick] (a2) arc (-180:0:.3cm);
	\node at ($ (a2) + (.3,-.05) $) {$\xi$};
	\node at ($ (a2) + (.4,.55) $) {\scriptsize{$b$}};
 \draw[thick, \Zstring, decorate, decoration={snake, segment length=1mm, amplitude=.2mm}] ($ (a2) + (.3,-.3)$) --node[left]{$\scriptstyle K$} ($ (a2) + (.3,-.7)$);
	\pairOfPantsTwo{(b1)}{}
	\draw[thick, blue] (c1) -- (1,1.9);
	\draw[thick, blue] ($ (b1) + (.3,.1) $) .. controls ++(90:.5cm) and ++(225:.4cm) .. (c1);
	\draw[thick, blue] ($ (b2) + (.3,.1) $) .. controls ++(90:.5cm) and ++(-45:.4cm) .. (c1);
	\filldraw (c1) circle (.05cm);
	\node at ($ (c1) + (-.2,.05) $) {\scriptsize{$\psi$}};
	\node at ($ (c1) + (.15,.3) $) {\scriptsize{$c$}};
	\topCylinderTwo{(d1)}{}
	\draw[thick, \Cstring] (1,1.9) -- (1,2.9);
	\colorHalfDottedEllipseTwo{($ (e1) + (0,-.1) $)}{.3}{.1}{\CstringOne}
	\roundNbox{unshaded}{($ (e1) + (.3,0) $)}{.25}{-.1}{-.1}{\scriptsize{$f$}}
	\node at ($ (d1) + (.15,.8) $) {\scriptsize{$d$}};
	\node at ($ (e1) + (.-.15,0) $) {\scriptsize{$x$}};
}
=
\tikzmath[yscale=-1]{
	\coordinate (a1) at (0,0);
	\coordinate (a2) at (1.4,0);
	\coordinate (b1) at (0,.3);
	\coordinate (b2) at (1.4,.3);
	\coordinate (c1) at (1,1.3);
	\coordinate (d1) at (.7,1.8);
	\coordinate (e1) at (.7,2.25);
	\emptyCylinder{(a1)}{.3}{.3}
	\draw[thick] (a1) arc (-180:0:.3cm);
	\node at ($ (a1) + (.3,-.05) $) {$\eta$};
 \draw[thick, \Zstring, decorate, decoration={snake, segment length=1mm, amplitude=.2mm}] ($ (a1) + (.3,-.3)$) --node[left]{$\scriptstyle H$} ($ (a1) + (.3,-.7)$);
	\emptyCylinder{(a2)}{.3}{.3}
	\draw[thick] (a2) arc (-180:0:.3cm);
	\node at ($ (a2) + (.3,-.05) $) {$\xi$};
 \draw[thick, \Zstring, decorate, decoration={snake, segment length=1mm, amplitude=.2mm}] ($ (a2) + (.3,-.3)$) --node[left]{$\scriptstyle K$} ($ (a2) + (.3,-.7)$);
	\pairOfPantsTwo{(b1)}{}
	\draw[thick, \Cstring] (c1) -- ($ (e1) + (.3,0) $) ;
	\draw[thick, blue] ($ (b1) + (.3,.1) $) .. controls ++(90:.4cm) and ++(225:.4cm) .. (c1);
	\draw[thick, blue] ($ (b2) + (.3,.1) $) .. controls ++(90:.4cm) and ++(-45:.4cm) .. (c1);
	\filldraw (c1) circle (.05cm);
	\topCylinderTwo{(d1)}{}
	\draw[thick, \Cstring] ($ (e1) + (.3,0) $) -- (1,2.9);
	\draw[thick, \CstringOne] ($ (e1) + (.15,0) $) .. controls ++(180:.1cm) and ++(90:.4cm) .. ($ (c1) + (-.23,.3) $) .. controls ++(270:.4cm) and ++(45:.4cm) .. ($ (a1) + (.01, .4) $); 
	\draw[thick, \CstringOne, dotted] ($ (a1) + (.01, .4) $) arc (180:0:.3cm and .1cm);
	\draw[thick, \CstringOne] ($ (e1) + (.45,0) $) .. controls ++(0:.1cm) and ++(90:.4cm) .. ($ (c1) + (.23,.3) $) .. controls ++(270:.4cm) and ++(135:.4cm) .. ($ (a2) + (.61, .4) $); 
	\draw[thick, \CstringOne, dotted] ($ (a2) + (.01, .4) $) arc (180:0:.3cm and .1cm);
	\draw[thick, \CstringOne] ($ (a1) + (.61,.4) $) .. controls ++(135:.4cm) and ++(180:.3cm) .. ($ (c1) + (0,-.25) $) ;
	\draw[thick, \CstringOne] ($ (a2) + (.01,.4) $) .. controls ++(45:.4cm) and ++(0:.3cm) .. ($ (c1) + (0,-.25) $) ;
	\roundNbox{unshaded}{($ (e1) + (.3,0) $)}{.25}{-.1}{-.1}{\scriptsize{$f$}}
}
=
\tikzmath[yscale=-1]{
\coordinate (z1) at (0,-1);
	\coordinate (z2) at (1.4,-1);
	\coordinate (a1) at (0,0);
	\coordinate (a2) at (1.4,0);
	\coordinate (b1) at (0,.3);
	\coordinate (b2) at (1.4,.3);
	\coordinate (c1) at (1,1.3);
	\coordinate (d1) at (.7,1.8);
	\coordinate (e1) at (.7,2.25);
	\emptyCylinder{(z1)}{.3}{.3}
	\draw[thick] (z1) arc (-180:0:.3cm);
	\node at ($ (z1) + (.3,-.05) $) {$\eta$};
	\bottomCylinderTwo{($ (z1) + (0,.3) $)}{.3}{1}
 \draw[thick, \Zstring, decorate, decoration={snake, segment length=1mm, amplitude=.2mm}] ($ (z1) + (.3,-.3)$) --node[left]{$\scriptstyle H$} ($ (z1) + (.3,-.7)$);
	\emptyCylinder{(z2)}{.3}{.3}
	\draw[thick] (z2) arc (-180:0:.3cm);
	\node at ($ (z2) + (.3,-.05) $) {$\xi$};
	\bottomCylinderTwo{($ (z2) + (0,.3) $)}{.3}{1}
 \draw[thick, \Zstring, decorate, decoration={snake, segment length=1mm, amplitude=.2mm}] ($ (z2) + (.3,-.3)$) --node[left]{$\scriptstyle K$} ($ (z2) + (.3,-.7)$);
	\pairOfPantsTwo{(b1)}{}
	\draw[thick, \Cstring] (c1) -- ($ (e1) + (.3,0) $) ;
	\draw[thick, \Cstring] ($ (c1) + (-.35,-.3) $) -- (c1);
	\draw[thick, \Cstring] ($ (z1) + (.3,.9) $) -- ($ (z1) + (.3,.4) $);
	\draw[thick, \Cstring] ($ (z1) + (.3,.9) $)  -- ($ (b1) + (.3,-.1) $) .. controls ++(90:.4cm) and ++(225:.4cm) .. ($ (c1) + (-.35,-.3) $);
	\draw[thick, \Cstring] ($ (c1) + (.35,-.3) $) -- (c1);
	\draw[thick, \Cstring] ($ (z2) + (.3,.9) $) -- ($ (z2) + (.3,.4) $);
	\draw[thick, \Cstring] ($ (z2) + (.3,.9) $) -- ($ (b2) + (.3,-.1) $) .. controls ++(90:.4cm) and ++(-45:.4cm) .. ($ (c1) + (.35,-.3) $);
	\filldraw (c1) circle (.05cm);
	\node[\Cstring] at ($ (z2) + (.4,1.6) $) {$\scriptstyle f$};
	\node[\Cstring] at ($ (z1) + (.2,1.6) $) {$\scriptstyle e$};
	\topCylinderTwo{(d1)}{}
	\draw[thick, \Cstring] ($ (e1) + (.3,0) $) -- (1,2.9);
	\draw[thick, \CstringOne] ($ (e1) + (.15,0) $) .. controls ++(180:.1cm) and ++(90:.4cm) .. ($ (c1) + (-.23,.4) $) .. controls ++(270:.4cm) and ++(60:.2cm) .. ($ (c1) + (-.35, -.3) $); 
	\draw[thick, \CstringOne] ($ (e1) + (.45,0) $) .. controls ++(0:.1cm) and ++(90:.4cm) .. ($ (c1) + (.23,.4) $) .. controls ++(270:.4cm) and ++(120:.2cm) .. ($ (c1) + (.35, -.3) $); 
	\draw[thick, \CstringOne] ($ (c1) + (-.35,-.3) $) .. controls ++(30:.3cm) and ++(150:.3cm) .. ($ (c1) + (.35,-.3) $); 
	\draw[thick, \CstringOne] ($ (z1) + (.3,.9) $) .. controls ++(210:.1cm) and ++(45:.3cm) .. ($ (z1) + (0,.5) $);
	\draw[thick, \CstringOne] ($ (z1) + (.3,.9) $) .. controls ++(-30:.1cm) and ++(135:.3cm) .. ($ (z1) + (.6,.5) $);
	\draw[thick, \CstringOne, dotted] ($ (z1) + (0,.5) $) arc (180:0:.3cm and .1cm);
	\draw[thick, \CstringOne] ($ (z2) + (.3,.9) $) .. controls ++(210:.1cm) and ++(45:.3cm) .. ($ (z2) + (0,.5) $);
	\draw[thick, \CstringOne] ($ (z2) + (.3,.9) $) .. controls ++(-30:.1cm) and ++(135:.3cm) .. ($ (z2) + (.6,.5) $);
	\draw[thick, \CstringOne, dotted] ($ (z2) + (0,0.5) $) arc (180:0:.3cm and .1cm);
	\roundNbox{unshaded}{($ (e1) + (.3,0) $)}{.25}{-.1}{-.1}{\scriptsize{$f$}}
	\filldraw ($ (c1) + (-.35,-.3) $) node[left]{$\scriptstyle \alpha$} circle (.05cm);
	\filldraw ($ (c1) + (.35,-.3) $) node[right]{$\scriptstyle 
	\beta$} circle (.05cm);
	\filldraw ($ (z1) + (.3,.9) $) node[right, xshift=-.1cm]{$\scriptstyle \alpha^\dag$} circle (.05cm);
	\filldraw ($ (z2) + (.3,.9) $) node[right, xshift=-.1cm, yshift=-.05cm]{$\scriptstyle \beta^\dag$} circle (.05cm);
}
\]
In the diagram on the right above,
we suppress a summation over simples $e,f\in\Irr(\cC)$ and orthonormal bases
$\{\alpha\}\subset\cC(x\otimes a\otimes \overline{x}\to e)$
and
$\{\beta\}\subset\cC(x\otimes a\otimes \overline{x}\to e)$.
Here, we have used the fusion relation \eqref{eq:FusionRelation} in $\cC$ to factor 
$$
\id_{x\otimes a\otimes \overline{x}}
=
\sum_{e\in\Irr(\cC)}\sum_{\beta\in\ONB(e\to x\otimes a\otimes \overline{x})} \alpha^\dag\alpha
$$
and similarly for $\id_{x\otimes b\otimes \overline{x}}$ through simples $f$.
Now we act the tube algebra on $\eta$ and $\xi$ on the top, and decompose the morphism on the bottom to get a linear combination of diagrams in the desired form.

The associator for $\Mod(\Tube(\cC))$ uses the associator in $\cC$ as usual for the Day convolution product.
We refer the reader to \cite[\S2.4-2.6]{MR3687214} for details.
\end{defn}

\section{Ungauging the Levin-Wen model}
\label{sec:UngaugingLevinWen}
\subsection{The local on-site Hilbert space}
\label{sec:UngaugingHilbertSpace}

Similar to the definition of our local 4-valent Hilbert space $\cH_v$ in \S\ref{sec:LWStringNet} above, we denote the space $\cC(X\otimes X\to X)$
by a single red trivalent vertex.
$$
\tikzmath{
\draw[dotted, rounded corners=5pt] (0,0) rectangle (.6,.6);
\fill[red] (.3,.3) circle (.05cm);
\draw[thick, red] (.3,.3) -- (.3,.6);
\draw[thick, red] (.3,.3) -- (.15,0);
\draw[thick, red] (.3,.3) -- (.45,0);
}
=
\cC(X\otimes X\to X)
$$
Recall that the canonical Lagrangian algebra in $Z(\cC)$ corresponding to $\cC$ as a gapped boundary to the vacuum is $A\coloneqq\Tr(1_\cC)$.

We define the local on-site Hilbert space for the ungauged theory by 
$$
\cK_v=
\tikzmath{
\draw[thick] (0,0) -- (0,1);
\draw[thick] (1,0) -- (1,1);
\halfDottedEllipse{(0,0)}{.5}{.2}
\halfDottedEllipse[thick, \Xstring]{(0,.5)}{.5}{.2}
\filldraw[thick, fill=white] (.5,1) ellipse (.5 and .2);
\fill[\Xstring] (.5,.3) circle (.05cm);
\draw[thick, red] (.5,.3) -- (.5,-.2);
}
=
\begin{aligned}
\\
Z(\cC)(\Tr(X)\to \Tr(1_\cC))
\\\cong
\cC(X\to F(A)).
\end{aligned}
$$
Each $\cK_v$ carries a canonical right $\Tube(\cC)$-action.
The corresponding object in $Z(\cC)$ is 
\begin{equation}
\label{eq:TubeRepForCanonicalLagrangian}
\bigoplus_{c\in\Irr(\cC)} (\cK_v)_c\otimes c
=
\bigoplus_{c\in\Irr(\cC)} \cC(c\to F(A))\otimes c
\cong 
A, 
\end{equation}
the canonical Lagrangian algebra.

The following local projector, which acts on the top of the tube corresponding to $\cK_v$, commutes with the $\Tube(\cC)$-action.
$$
B_v=
\frac{1}{D_\cC}
\sum_{c\in\Irr(\cC)}
d_c
\,
\tikzmath{
\draw[thick] (0,0) -- (0,1);
\draw[thick] (1,0) -- (1,1);
\halfDottedEllipse{(0,0)}{.5}{.2}
\halfDottedEllipse[thick, blue]{(0,.5)}{.5}{.2}
\node[blue] at (.5,.45) {$\scriptstyle c$};
\filldraw[thick, fill=white] (.5,1) ellipse (.5 and .2);
}
=
\tikzmath{
\draw[thick] (0,0) arc (180:0:.5 and .4);
\draw[thick] (0,1) arc (-180:0:.5 and .4);
\halfDottedEllipse{(0,0)}{.5}{.2}
\filldraw[thick, fill=white] (.5,1) ellipse (.5 and .2);
}
=
i^\dag\circ i
$$
That $B_v=i^\dag i$ was shown in \cite[Rem.~2.10]{2305.14068}.
In terms of string diagrams, $B_v$ caps off the tube corresponding to $\cK_v$, after which the $X$ strand may be isotoped over the top \cite[Lem.~3.12]{2301.11114}:
\begin{align*}
B_v\cK_v
&=
\tikzmath{
\draw[thick] (0,0) -- (0,1);
\draw[thick] (1,0) -- (1,1);
\halfDottedEllipse{(0,0)}{.5}{.2}
\draw[thick, dotted] (.5,1) ellipse (.5 and .2);
\draw[thick] (0,1) arc (180:0:.5 and .4);
\draw[thick] (0,2) arc (-180:0:.5 and .4);
\halfDottedEllipse[thick, red]{(0,.5)}{.5}{.2}
\filldraw[thick, fill=white] (.5,2) ellipse (.5 and .2);
\fill[red] (.5,.3) circle (.05cm);
\draw[thick, red] (.5,.3) -- (.5,-.2);
}
=
\tikzmath{
\draw[thick] (0,0) -- (0,1);
\draw[thick] (1,0) -- (1,1);
\halfDottedEllipse{(0,0)}{.5}{.2}
\draw[thick, dotted] (.5,1) ellipse (.5 and .2);
\draw[thick] (0,1) arc (180:0:.5 and .4);
\draw[thick] (0,2) arc (-180:0:.5 and .4);
\draw[thick, red] (.5,.5) circle (.2cm);
\filldraw[thick, fill=white] (.5,2) ellipse (.5 and .2);
\fill[red] (.5,.3) circle (.05cm);
\draw[thick, red] (.5,.3) -- (.5,-.2);
}
=
\bbC\cdot
\tikzmath{
\draw[thick] (0,0) arc (180:0:.5 and .4);
\draw[thick] (0,1) arc (-180:0:.5 and .4);
\halfDottedEllipse{(0,0)}{.5}{.2}
\filldraw[thick, fill=white] (.5,1) ellipse (.5 and .2);
}\,.
\intertext{We can also directly calculate}
B_v\cK_v
&=
Z(\cC)(\Tr(X)\to 1_{Z(\cC)})
\cong
\cC(X\to 1_\cC)
\cong \bbC.
\end{align*}
Hence the projector $B_v$ is minimal, with a one-dimensional image.

\subsection{The tensor product space of tube algebra modules}
\label{sec:SpineCoralSkeinModule}

Similar to the gauging story for $G$-symmetry protected topological phases ($G$-SPTs),
we have a unitary isomorphism from an invariant subspace of the ungauged model to the image of the projector $P_A$ on the Levin-Wen Hilbert space $\cH_{\rm tot}=\bigotimes_v \cH_v$.
We use two ingredients to write down this unitary: the mathematical object which plays the role of the group $G$ for our ungauged model, and the notion of the invariant subspace.
We now look more closely at the ingredients in the $G$-SPT case in order to generalize to the ungauged Levin-Wen setting.

The finite dimensional Hilbert space representations of a group $G$ form the symmetric unitary tensor category $\Rep(G)$, which comes equipped with a symmetric fiber functor to $\Hilb$.
This symmetric tensor product allows us to take an \emph{unordered} tensor product of group representations, which we may view as spread on the sites of a 2D lattice.
The unordering means that given any ordering $x_1,\cdots, x_n$ and any permutation $x_{\sigma(1)},\cdots, x_{\sigma(n)}$, there is a unique element of the symmetric group $\fS_n$ taking one ordering to the other, and these orderings compose in the obvious way.
This element of $\fS_n$ gives a canonical unitary in $\Rep(G)$ taking one tensor product ordering to the other.

The existence of the fiber functor $\Rep(G)\to \Hilb$ endows the group algebra $\bbC[G]$ with the structure of a Hopf algebra.
In particular, the coproduct of $\bbC[G]$ allows a boost of the local $G$-symmetry to a global $G$-symmetry acting at every site simultaneously.

For the ungauged model, the role of the local $G$-symmetry is played by the tube algebra $\Tube(\cC)$ which acts locally on each site on the bottom of the tube corresponding to $\cK_v$.
The tube algebra is not a Hopf algebra, but its category of right representations $\Mod(\Tube(\cC))$ is equivalent to the modular fusion category $Z(\cC)$.
As modular is the opposite of symmetric for braided fusion categories, we need a generalization of the notion of unordered tensor product in order to place our local Hilbert spaces $\cK_v$ at sites on a 2D lattice.
Fortunately, a modular fusion category is spherical braided, and in particular, it is \emph{balanced}.\footnote{A \emph{balancing} on a braided tensor category is a natural isomorphism $\theta: \id_\cC\Rightarrow \id_\cC$ with $(\theta_a\otimes \theta_b) \circ \beta_{b,a}\circ \beta_{a,b}=\theta_{a\otimes b}$.}
In such a category, it makes sense to take a tensor product of $n$ objects located at $n$ points in the 2D plane equipped with tangent vectors \cite[Prop.~7.6]{MR1989873}.\footnote{The article \cite{MR1989873} uses the term ribbon in place of balanced.}
Given another configuration of $n$ points equipped with tangent vectors in the 2D plane, there is a unique element of the ribbon braid group $\cR\cB_n=\bbZ^n \rtimes \cB_n$ \cite[\S5]{MR1989873} \cite[Def.~4.8]{MR4528312} taking one configuration to the other.
Indeed, $\cR\cB_n$ is the fundamental group of the configuration space of $n$ unordered oriented points equipped with tangent vectors in the 2D plane.
This element of $\cR\cB_n$ gives a canonical unitary isomorphism in $Z(\cC)$ from one tensor product ordering to the other.

We now view our local Hilbert spaces $\cK_v$ at sites on a 2D lattice, namely on the \emph{plaquettes} of the 2D lattice on which our Levin-Wen model lives.
To boost the local on-site $\Tube(\cC)$ action to a global $\Tube(\cC)$ action, we take the ribbon-braided tensor product of the objects $\cK_v$ in the modular fusion category $\Mod(\Tube(\cC))\cong Z(\cC)$.
By the preceding paragraph, it makes sense to take such a tensor product of objects located at these sites on our 2D lattice, where by convention, the tangent vector at each site points upwards.
This tensor product is the iterated Day convolution product defined in \S\ref{sec:TubeAlgebra}, and yields a tube which splits into many tubes as it grows upwards, like a coral attached to the sea floor as in Figure \ref{fig:SpineCoral} below.
Our coral has red $X$-strands and trivalent vertices corresponding to the Hilbert space $\cC(X\to X\otimes X)$ with the skein module inner product,
and a basis for this space is the so-called \emph{spine basis}.
We refer to the object in $\Mod(\Tube(\cC))$  obtained in this way as the \emph{spine coral skein module}.

\begin{widetext}
\begin{center}
\begin{figure}[!hb]
$$
\tikzmath[scale=1]{
\coordinate (a) at (1.5,-3);
\coordinate (b11) at (-.5,0);
\coordinate (b21) at (1.5,0);
\coordinate (b31) at (3.5,0);
\coordinate (b12) at (-.25,2);
\coordinate (b22) at (1.75,2);
\coordinate (b32) at (3.75,2);
\coordinate (b13) at (0,4);
\coordinate (b23) at (2,4);
\coordinate (b33) at (4,4);
\halfDottedEllipse{(a)}{.5}{.2}
\draw[thick] (b12) -- ($ (b12) + (0,-.5) $) .. controls ++(270:.8cm) and ++(90:.8cm) .. ($ (b11) + (.6,0) $) ;
\draw[thick] (b13) -- ($ (b13) + (0,-.5) $) .. controls ++(270:.8cm) and ++(90:.8cm) .. ($ (b12) + (.6,0) $) ;
\draw[thick] ($ (b11) + (1,0) $) -- ($ (b11) + (1,-.5) $) .. controls ++(270:.3cm) and ++(110:.3cm) .. ($ (b11) + (1.3,-1.2) $) ;
\draw[thick] (b21) -- ($ (b21) + (0,-.5) $) .. controls ++(270:.3cm) and ++(70:.3cm) .. ($ (b21) + (-.15,-1.2) $) ;
\draw[thick] ($ (b21) + (1,0) $) -- ($ (b21) + (1,-.5) $) .. controls ++(270:.3cm) and ++(110:.3cm) .. ($ (b21) + (1.15,-1.2) $) ;
\draw[thick] ($ (b31) + (1,0) $) -- ($ (b31) + (1,-.5) $) .. controls ++(270:.3cm) and ++(70:.3cm) .. ($ (b31) + (.7,-1.2) $) ;
\draw[thick] (b31) -- ($ (b31) + (0,-.5) $) .. controls ++(270:.3cm) and ++(70:.3cm) .. ($ (b31) + (-.3,-1.2) $) ;
\draw[thick] ($ (b12) + (1,0) $) -- ($ (b12) + (1,-.5) $) .. controls ++(270:.3cm) and ++(110:.3cm) .. ($ (b12) + (1.3,-1.2) $) ;
\draw[thick] (b22) -- ($ (b22) + (0,-.5) $) .. controls ++(270:.3cm) and ++(70:.3cm) .. ($ (b22) + (-.15,-1.2) $) ;
\draw[thick] ($ (b22) + (1,0) $) -- ($ (b22) + (1,-.5) $) .. controls ++(270:.3cm) and ++(110:.3cm) .. ($ (b22) + (1.15,-1.2) $) ;
\draw[thick] ($ (b32) + (1,0) $) -- ($ (b32) + (1,-.5) $) .. controls ++(270:.3cm) and ++(70:.3cm) .. ($ (b32) + (.7,-1.2) $) ;
\draw[thick] (b32) -- ($ (b32) + (0,-.5) $) .. controls ++(270:.3cm) and ++(70:.3cm) .. ($ (b32) + (-.3,-1.2) $) ;
\draw[thick] (a) .. controls ++(90:.8cm) and ++(270:1.2cm) .. ($ (b11) + (0,-.5) $) -- (b11);
\draw[thick] ($ (a) + (1,0) $) .. controls ++(90:.8cm) and ++(270:1.2cm) .. (5,-1) -- ($ (b33) + (1,0) $);
\draw[thick] ($ (b13) + (1,0) $) -- ($ (b13) + (1,-.5) $) arc (-180:0:.5) -- (b23);
\draw[thick] ($ (b23) + (1,0) $) -- ($ (b23) + (1,-.5) $) arc (-180:0:.5) -- (b33);
\foreach \x in {0,2,4}{
\foreach \y in {0,2,4}{
\halfDottedEllipse[thick, red]{(\x+\y*.125-.5,\y-.5)}{.5}{.2}
\fill[red] {(\x+\y*.125,\y-.7)} circle (.05cm);
}}
\draw[thick, red] ($ (b11) + (.5,-.7) $)  .. controls ++(270:.8cm) and ++(90:.8cm) .. ($ (a) + (.5,-.2) $);
\draw[thick, red] ($ (b21) + (.5,-.7) $)  .. controls ++(270:.8cm) and ++(45:.6cm) .. (1.19,-2.11); 
\fill[red] (1.19,-2.11) circle (.05cm);
\draw[thick, red] ($ (b31) + (.5,-.7) $)  .. controls ++(270:.8cm) and ++(45:.6cm) .. (1.78,-2.69); 
\fill[red] (1.78,-2.69) circle (.05cm);
\draw[thick, red] ($ (b12) + (.5,-.7) $)  .. controls ++(270:.8cm) and ++(90:.8cm) .. ($ (b11) + (1.5,-.7) $) .. controls ++(270:.6cm) and ++(45:.4cm) .. (.6,-1.6); 
\fill[red] (.6,-1.6) circle (.05cm);
\draw[thick, red] ($ (b22) + (.5,-.7) $)  .. controls ++(270:.8cm) and ++(90:.8cm) .. ($ (b21) + (1.5,-.7) $) .. controls ++(270:.8cm) and ++(20:.2cm) .. (1.72,-1.6);  
\fill[red] (1.72,-1.6) circle (.05cm);
\draw[thick, red] ($ (b32) + (.5,-.7) $)  .. controls ++(270:.8cm) and ++(90:.8cm) .. ($ (b31) + (1.3,-.7) $) .. controls ++(270:.8cm) and ++(20:.2cm) .. (3.37,-1.6);  
\fill[red] (3.4,-1.58) circle (.05cm);
\draw[thick, red] ($ (b13) + (.5,-.7) $)  .. controls ++(270:.8cm) and ++(90:.8cm) .. ($ (b12) + (1.5,-.7) $) .. controls ++(270:.6cm) and ++(45:.4cm) .. (.8,0); 
\fill[red] (.8,0) circle (.05cm);
\draw[thick, red] ($ (b23) + (.5,-.7) $)  .. controls ++(270:.8cm) and ++(90:.8cm) .. ($ (b22) + (1.5,-.7) $) .. controls ++(270:.6cm) and ++(45:.4cm) .. (2.8,0); 
\fill[red] (2.8,0) circle (.05cm);
\draw[thick, red] ($ (b33) + (.5,-.7) $)  .. controls ++(270:.6cm) and ++(90:1.2cm) .. ($ (b32) + (1.15,-.7) $) .. controls ++(270:.6cm) and ++(45:.2cm) .. (4.65,0); 
\fill[red] (4.65,0) circle (.05cm);
\foreach \x in {0,2,4}{
\foreach \y in {0,2,4}{
\filldraw[thick, fill=white] (\x+\y*.125,\y) ellipse (.5 and .2);
}}
}
\qquad\qquad
\raisebox{-.5\height}{\includegraphics[scale=.75]{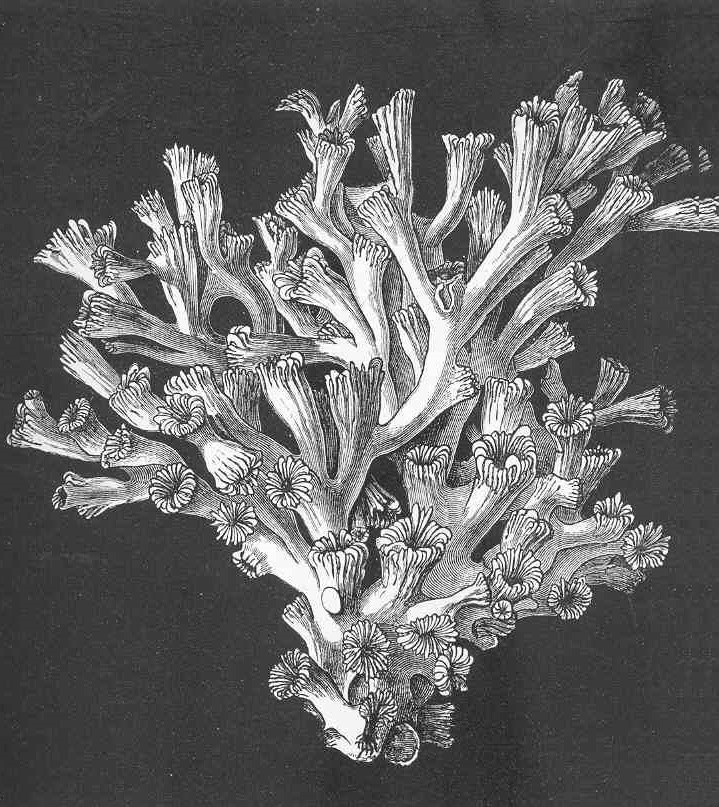}}
$$    
\caption{Left: The spine coral skein module produced by an iterated Day convolution product. 
\\
Right: A sketch of \textit{ 
Lophohelia prolifera, Pallas (sp.)} \cite{CoralImage} }
    \label{fig:SpineCoral}
\end{figure}
\end{center}
\end{widetext}

Under the equivalence $\Mod(\Tube(\cC))\cong Z(\cC)$, 
since $\cK_v$ corresponds to the canonical Lagrangian $A$ by \eqref{eq:TubeRepForCanonicalLagrangian},
the spine coral skein module corresponds exactly to the ribbon-braided tensor product of a copy of the canonical Lagrangian algebra $A$ located at each plaquette of our 2D lattice.
We denote this object by $A^{\otimes p\in\Lambda}$.
As a right module of $\Tube(\cC)$,
the Hilbert space of the spine coral skein module is
\begin{equation}
\label{eq:SCSMHilbertSpace}
Z(\cC)(\Tr(X)\to A^{\otimes p\in\Lambda}).
\end{equation}

\begin{rem}
Unlike most lattice models of `spins,' our Hilbert space \eqref{eq:SCSMHilbertSpace} is not a tensor product space over the Hilbert spaces on sites.
In \S\ref{sec:LocalOperators}, we show that, nonetheless, there is a well-defined notion of local operators.
\end{rem}

\subsection{The invariant subspace and the gauging map}
\label{sec:TheGaugingMap}

Observe that the spine coral skein module carries a right $\Tube(\cC)$-action.
To get the invariant subspace, we take the maps from the trivial $\Tube(\cC)$-representation, namely $1_{Z(\cC)}$, to the spine coral skein module.
To do so, one first projects to the subspace where the boundary label is $1_\cC$ and then applies the cup.
For the spine coral skein module appearing in Figure \ref{fig:SpineCoral}, the corresponding invariant subspace is given by the left hand side of Figure \ref{fig:GaugingMap} below.

Topologically, the invariant subspace is equivalent to the skein module of a sphere with $\ell^2$ punctures, if we had started with an $\ell\times \ell$ 2D lattice of sites supporting our local Hilbert space $\cK_v$.
We may view this punctured sphere as a punctured 2D plane with a point at $\infty$.
Since the spine basis of our punctured skein module is bounded away from $\infty$ (with the caveat that we may still perform spherical isotopy for string diagrams around the outside of the lattice), 
we may draw our spine basis on the punctured 2D plane as follows.

\begin{widetext}
\begin{center}
\begin{figure}[!ht]
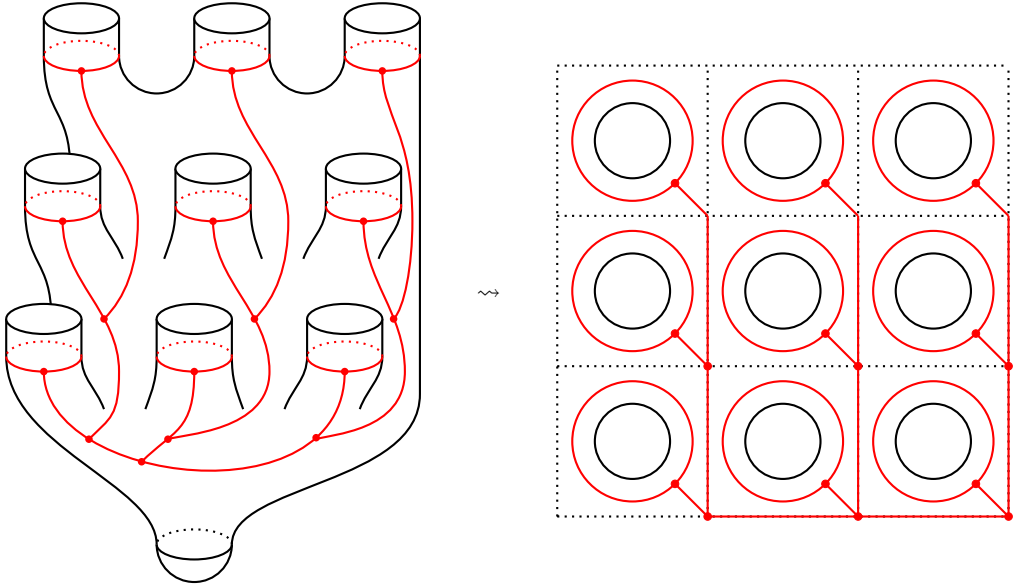

$$
\tikzmath{
\coordinate (a) at (1.5,-3);
\coordinate (b11) at (-.5,0);
\coordinate (b21) at (1.5,0);
\coordinate (b31) at (3.5,0);
\coordinate (b12) at (-.25,2);
\coordinate (b22) at (1.75,2);
\coordinate (b32) at (3.75,2);
\coordinate (b13) at (0,4);
\coordinate (b23) at (2,4);
\coordinate (b33) at (4,4);
\halfDottedEllipse{(a)}{.5}{.2}
\draw[thick] (a) arc(-180:0:.5cm); 
\draw[thick] (b12) -- ($ (b12) + (0,-.5) $) .. controls ++(270:.8cm) and ++(90:.8cm) .. ($ (b11) + (.6,0) $) ;
\draw[thick] (b13) -- ($ (b13) + (0,-.5) $) .. controls ++(270:.8cm) and ++(90:.8cm) .. ($ (b12) + (.6,0) $) ;
\draw[thick] ($ (b11) + (1,0) $) -- ($ (b11) + (1,-.5) $) .. controls ++(270:.3cm) and ++(110:.3cm) .. ($ (b11) + (1.3,-1.2) $) ;
\draw[thick] (b21) -- ($ (b21) + (0,-.5) $) .. controls ++(270:.3cm) and ++(70:.3cm) .. ($ (b21) + (-.15,-1.2) $) ;
\draw[thick] ($ (b21) + (1,0) $) -- ($ (b21) + (1,-.5) $) .. controls ++(270:.3cm) and ++(110:.3cm) .. ($ (b21) + (1.15,-1.2) $) ;
\draw[thick] ($ (b31) + (1,0) $) -- ($ (b31) + (1,-.5) $) .. controls ++(270:.3cm) and ++(70:.3cm) .. ($ (b31) + (.7,-1.2) $) ;
\draw[thick] (b31) -- ($ (b31) + (0,-.5) $) .. controls ++(270:.3cm) and ++(70:.3cm) .. ($ (b31) + (-.3,-1.2) $) ;
\draw[thick] ($ (b12) + (1,0) $) -- ($ (b12) + (1,-.5) $) .. controls ++(270:.3cm) and ++(110:.3cm) .. ($ (b12) + (1.3,-1.2) $) ;
\draw[thick] (b22) -- ($ (b22) + (0,-.5) $) .. controls ++(270:.3cm) and ++(70:.3cm) .. ($ (b22) + (-.15,-1.2) $) ;
\draw[thick] ($ (b22) + (1,0) $) -- ($ (b22) + (1,-.5) $) .. controls ++(270:.3cm) and ++(110:.3cm) .. ($ (b22) + (1.15,-1.2) $) ;
\draw[thick] ($ (b32) + (1,0) $) -- ($ (b32) + (1,-.5) $) .. controls ++(270:.3cm) and ++(70:.3cm) .. ($ (b32) + (.7,-1.2) $) ;
\draw[thick] (b32) -- ($ (b32) + (0,-.5) $) .. controls ++(270:.3cm) and ++(70:.3cm) .. ($ (b32) + (-.3,-1.2) $) ;
\draw[thick] (a) .. controls ++(90:.8cm) and ++(270:1.2cm) .. ($ (b11) + (0,-.5) $) -- (b11);
\draw[thick] ($ (a) + (1,0) $) .. controls ++(90:.8cm) and ++(270:1.2cm) .. (5,-1) -- ($ (b33) + (1,0) $);
\draw[thick] ($ (b13) + (1,0) $) -- ($ (b13) + (1,-.5) $) arc (-180:0:.5) -- (b23);
\draw[thick] ($ (b23) + (1,0) $) -- ($ (b23) + (1,-.5) $) arc (-180:0:.5) -- (b33);
\foreach \x in {0,2,4}{
\foreach \y in {0,2,4}{
\halfDottedEllipse[thick, red]{(\x+\y*.125-.5,\y-.5)}{.5}{.2}
\fill[red] {(\x+\y*.125,\y-.7)} circle (.05cm);
}}
\draw[thick, red] ($ (b11) + (.5,-.7) $)  .. controls ++(270:1.5cm) and ++(270:2cm) .. ($ (b31) + (.5,-.7) $);
\draw[thick, red] ($ (b21) + (.5,-.7) $)  .. controls ++(270:.8cm) and ++(45:.6cm) .. (1.3,-1.9); 
\fill[red] (1.3,-1.9) circle (.05cm);
\draw[thick, red] ($ (b12) + (.5,-.7) $)  .. controls ++(270:.8cm) and ++(90:.8cm) .. ($ (b11) + (1.5,-.7) $) .. controls ++(270:.6cm) and ++(45:.4cm) .. (.6,-1.6); 
\fill[red] (.6,-1.6) circle (.05cm);
\draw[thick, red] ($ (b22) + (.5,-.7) $)  .. controls ++(270:.8cm) and ++(90:.8cm) .. ($ (b21) + (1.5,-.7) $) .. controls ++(270:.8cm) and ++(20:.2cm) .. (1.65,-1.6);  
\fill[red] (1.65,-1.6) circle (.05cm);
\draw[thick, red] ($ (b32) + (.5,-.7) $)  .. controls ++(270:.8cm) and ++(90:.8cm) .. ($ (b31) + (1.3,-.7) $) .. controls ++(270:.8cm) and ++(20:.2cm) .. (3.62,-1.6);  
\fill[red] (3.62,-1.58) circle (.05cm);
\draw[thick, red] ($ (b13) + (.5,-.7) $)  .. controls ++(270:.8cm) and ++(90:.8cm) .. ($ (b12) + (1.5,-.7) $) .. controls ++(270:.6cm) and ++(45:.4cm) .. (.8,0); 
\fill[red] (.8,0) circle (.05cm);
\draw[thick, red] ($ (b23) + (.5,-.7) $)  .. controls ++(270:.8cm) and ++(90:.8cm) .. ($ (b22) + (1.5,-.7) $) .. controls ++(270:.6cm) and ++(45:.4cm) .. (2.8,0); 
\fill[red] (2.8,0) circle (.05cm);
\draw[thick, red] ($ (b33) + (.5,-.7) $)  .. controls ++(270:.6cm) and ++(90:1.2cm) .. ($ (b32) + (1.15,-.7) $) .. controls ++(270:.6cm) and ++(45:.2cm) .. (4.65,0); 
\fill[red] (4.65,0) circle (.05cm);
\foreach \x in {0,2,4}{
\foreach \y in {0,2,4}{
\filldraw[thick, fill=white] (\x+\y*.125,\y) ellipse (.5 and .2);
}}
}
\qquad
\rightsquigarrow
\qquad
\tikzmath{
\draw[dotted,thick,scale=2,xshift=.5cm,yshift=.5cm] (0,0) grid (3,3);
\foreach \x in {1,2,3}{
\foreach \y in {1,2,3}{
\coordinate (b\x\y) at ($ \x*(2,0)+ \y*(0,2) $);
\draw[thick] ($ \x*(2,0)+ \y*(0,2) $) circle (.5cm);
\draw[thick,red] ($ \x*(2,0)+ \y*(0,2) $) circle (.8cm);
\filldraw[red] ($ (b\x\y) + (-45:.8cm) $) circle (.05cm);
\draw[thick, red] ($ (b\x\y) + (-45:.8cm) $) -- ($ (b\x\y) + (1,-1) $);
}}
\foreach \x in {1,2,3}{
\draw[thick, red] ($ (b\x1) + (1,-1) $) -- ($ (b\x3) + (1,-1) $);
\filldraw[red] ($ (b\x1) + (1,-1) $) circle (.05cm);
\filldraw[red] ($ (b\x2) + (1,-1) $) circle (.05cm);
}
\draw[thick, red] ($ (b11) + (1,-1) $) -- ($ (b31) + (1,-1) $);
}
$$
\caption{\label{fig:GaugingMap}$\Tube(\cC)$-invariant subspace and the gauging map.
The Levin-Wen lattice is depicted by the dotted grid.}
\end{figure}
\end{center}
\end{widetext}

We then get a unitary isomorphism to the image of $P_A$ in the Levin-Wen microscopic Hilbert space, modulo spherical isotopy (see Remark \ref{rem:SphericalIsotopy} below), by resolving the spine basis elements into the local vertex Hilbert spaces $\cH_v$ which lie equidistant from 4 neighboring punctures.
On the right hand side of Figure \ref{fig:GaugingMap}, the Levin-Wen lattice is depicted by the dotted grid, and the vertices of this grid support the vertex spaces $\cH_v$.
The spaces on the boundary are modified to have only 2 or 3 incoming edges, which can be done by forcing the `ghost edge' labels past the boundary to be $1_\cC$.

\begin{rem}
\label{rem:SphericalIsotopy}
There is a subtlety here about spherical isotopy around the outside of the punctured 2D skein module, which is due to the fact that this is really the skein module of a punctured sphere.
Generally speaking, in the Levin-Wen local Hilbert space, spherical isotopy around the outside of the lattice is not allowed.
However, if we draw our lattice on a contractible region on a sphere, then sphericality of $\cC$ implies spherical isotopy around the outside is allowed in the skein module. Specifically, if the `ghost edges' of our finite lattice extending outside the contractible region are labelled by the object $1_\cC$, then \cite[Prop.~2.6]{2305.14068} identifies the image of $P_A$ on the finite lattice with 
$$
\im(P_A)\cong \cC(1_\cC \to F(A)^{\otimes p\in\Lambda}).
$$
\end{rem}

\begin{rem}
    In the ungauged model, there is a unique symmetric gapped ground state, even when the model is on a finite patch of lattice.
    In SPT theories protected by a global symmetry group, this would indicate a completely trivial state 
    which is equivalent to a product state.
    We take this as a physical indication that our state is trivial, even though our underlying Hilbert space is not a tensor product Hilbert space.
    Therefore, we interpret our gauging procedure as starting with a trivial theory which undergoes twisted gauging to produce topological order. 
    \end{rem}

\begin{rem}

When gauging a global on-site symmetry of a lattice theory, it is standard to start with a theory defined on a tensor product Hilbert space with the symmetry acting as a tensor product of operators on each site.
That is, if the global symmetry group is $G$, then each site is a $\mathbb{C}[G]$-module and the global symmetry acts as the tensor product over all sites.
In our case, the global symmetry is the tube algebra $\Tube(\cC)$, and the ordinary tensor product of modules of $\Tube(\cC)$ is not, in general, a module of $\Tube(\cC)$.
This is why our ungauged theory must be defined on a Hilbert space which is not a tensor product over sites.

As we see in Section  \S\ref{sec:LocalOperators} below, although the structure of the Hilbert space is non-local in some sense, there is a well-defined local structure on the operators, and therefore observables, of the theory.
In fact, the $\Tube(\cC)$ symmetry still allows for symmetric local operators with finite support in every local patch of space, unlike other non-on-site symmetries such as translation or inversion symmetry. 

Alternatively, if one insists on a true tensor product Hilbert space, one may construct a theory where the Hilbert space is a tensor product of $\mathcal{K}_v$ on each site $v$, together with some ancilla.
The Hilbert space in our previous construction is a subspace of this tensor product space and one may consider the $\Tube(\cC)$ symmetry as being emergent by placing an energetic cost on leaving this subspace.

\end{rem}
\subsection{Local operator algebras of observables on either side of the gauging map}
\label{sec:LocalOperators}

In this section, we provide an operator algebraic formalism for describing how the local symmetric operators of the ungauged theory translate to the local operators of the gauged theory, even on the infinite lattice where the Hilbert space is difficult to define in a natural way.
To be specific, we prove that our gauging procedure sends local equivariant operators to local equivariant operators. 

Recall that the gauged Hilbert space is the subspace of the Levin-Wen Hilbert space in the image of the $A_\ell$ terms.
The condition $A_\ell|\omega\rangle=|\omega\rangle$ on states $|\omega\rangle$ is called the \emph{gauge constraint}.
Now it is clear that each $B_v$ term in the Hamiltonian is mapped under conjugating by our gauging unitary to the corresponding $B_p$ term in the Levin-Wen Hamiltonian. 
Hence the state induced by the gauging map from the ground state of the ungauged theory is the ground state of the gauged theory, namely, the Levin-Wen ground state.
(Recall we started with a punctured sphere, and there is a unique Levin-Wen ground state on the sphere.)

As discussed in \S\ref{sec:SpineCoralSkeinModule}, 
under the equivalence $\Mod(\Tube(\cC))\cong Z(\cC)$,
$\cK_v$ corresponds to the canonical Lagrangian $A$
and
the spine coral skein module corresponds to $A^{\otimes p\in\Lambda}$, the ribbon-braided tensor product of $A$ located at the sites of the plaquettes of our 2D lattice.
By \eqref{eq:SCSMHilbertSpace},
the Hilbert space of the spine coral skein module as a $\Tube(\cC)$-module is
$$
Z(\cC)(\Tr(X)\to A^{\otimes p\in\Lambda}).
$$
Since $\Tube(\cC)\cong \End_{Z(\cC)}(\Tr(X))$ and $Z(\cC)(z\to\Tr(X))\neq 0$ for every $z\in \Irr(Z(\cC))$, by the Yoneda Lemma, the commutant of the right $\Tube(\cC)$-action on the spine coral skein module is 
\begin{equation}
\label{eq:Tube(C)Commutant}
\Tube(\cC)'=\End_{Z(\cC)}(A^{\otimes p\in\Lambda}).    
\end{equation}
We warn the reader that the commutant of the $\Tube(\cC)$-action on the spine coral skein module is not the space of all endomorphisms of the $\Tube(\cC)$-invariant subspace.
The first is an interesting multimatrix algebra, whereas the second is a full matrix algebra.

We now consider the local operator algebras in the thermodynamic limit on either side by looking at the inductive limit of grids of size $\ell\times \ell$ as $\ell\to\infty$.
Looking at a bounded contractible patch $\Lambda$ in the 2D lattice, 
\begin{equation}
\label{eq:SampleLambda}    
\tikzmath[scale=.5]{
\draw[step=1.0,red,thick] (0.5,0.5) grid (6.5,6.5);
\foreach \i in {1,2,...,6} {
\foreach \j in {1,2,...,6} {
        \filldraw[thick, red] (\i,\j)circle (.1cm);
}}
\draw[thick, rounded corners=5pt] (1.5,2.5) -- (1.5,4.5) -- (3.5,4.5) -- (3.5, 5.5) -- (5.5,5.5) -- (5.5,2.5) -- (3.5,2.5) -- (3.5,1.5) -- (1.5,1.5) -- (1.5,2.5);
\node at (3.5,3.5) {$\scriptstyle \Lambda$};
}
\end{equation}
we consider the local algebra of observables on $\Lambda$ on either side of the gauging unitary isomorphism.

On the ungauged side, we can perform a local isotopy of the spine coral skein module to consolidate the portion of the coral inside $\Lambda$.
This results in a Day convolution product between the spine coral skein module associated to sites in $\Lambda$ with the spine coral skein module associated to sites in $\Lambda^c$.
As in \eqref{eq:Tube(C)Commutant}, the algebra of observables localized in $\Lambda$ then corresponds to the commutant of the $\Tube(\cC)$-action on the spine coral skein module associated to $\Lambda$, i.e.,
$$
\fA(\Lambda)\coloneqq \End_{Z(\cC)}(A^{\otimes p\in\Lambda}).
$$

The local algebras $\fA(\Lambda)$ form a \emph{(local) net of algebras} in the sense of \cite{2307.12552} based on \cite[\S6.2]{MR1441540}.
Observe that whenever we have an inclusion of bounded contractible patches $\Lambda \subset \Gamma$, there is a canonical unital inclusion $\fA(\Lambda) \hookrightarrow \fA(\Gamma)$.
given by tensoring with $\id_A$ for every $A\in \Gamma\setminus \Lambda$.
If $\Lambda$ and $\Delta$ are disjoint bounded contractible patches, then in any bounded contractible patch $\Gamma$ containing $\Lambda\cup \Delta$, we have $[\fA(\Lambda),\fA(\Delta)]=0$ inside $\fA(\Gamma)$.

Under conjugation by the gauging unitary isomorphism to $\im(P_A)$ on the Levin-Wen side, we get a certain subalgebra of the local operators in $\bigotimes_{v\in\Lambda} \cB(\cH_v)$ which act on the local Hilbert space $\cH(\Lambda):=\bigotimes_{v\in\Lambda}\cH_v$.
By \eqref{eq:IntermediateSpace},
we have a unitary isomorphism 
\begin{align}
P_A\cH(\Lambda)
&\cong
\cC(X^{\otimes \partial \Lambda}\to F(A)^{\otimes p\in \Lambda})
\notag
\\&\cong
Z(\cC)(\Tr(X^{\otimes \partial \Lambda})\to A^{\otimes p\in \Lambda}).
\label{eq:IntermediateSpace2}
\end{align}
Observe that the space $P_A\cH(\Lambda)$
carries an action of an algebra Morita equivalent to $\Tube(\cC)$ called
$$
\Tube_\cC(\partial \Lambda)
=
\left\{\,
\tikzmath[scale=.5]{
\filldraw[thick, fill=white, rounded corners=5pt] (1.5,2.5) -- (1.5,4.5) -- (3.5,4.5) -- (3.5, 5.5) -- (5.5,5.5) -- (5.5,2.5) -- (3.5,2.5) -- (3.5,1.5) -- (1.5,1.5) -- (1.5,2.5);
\node at (3.5,3.5) {$\scriptstyle \Lambda$};
\draw[thick, rounded corners=5pt] (-.5,2.5) -- (-.5,6.5) -- (1.5,6.5) -- (1.5, 7.5) -- (7.5,7.5) -- (7.5,.5) -- (5.5,.5) -- (5.5,-.5) -- (-.5,-.5) -- (-.5,2.5);
\draw[name path=A--B, thick,red, rounded corners=5pt] (.5,2.5) -- (.5,5.5) -- (2.5,5.5) -- (2.5, 6.5) -- (6.5,6.5) -- (6.5,1.5) -- (4.5,1.5) -- (4.5,.5) -- (.5,.5) -- (.5,2.5);
\draw[name path=C1--D1, thick,red] (1.5,2) to[out=180, in=0] (-.5,1);
\draw[name path=C2--D2, thick,red] (1.5,3) to[out=180, in=0] (-.5,3);
\draw[name path=C3--D3,thick,red] (1.5,4) to[out=180, in=0] (-.5,5);
\draw[name path=C4--D4,thick,red] (2,4.5) to[out=90, in=-90] (0,6.5);
\draw[name path=C5--D5,thick,red] (3,4.5) to[out=90, in=-90] (1,6.5);
\draw[name path=C6--D6,thick,red] (3.5,5) to[out=180,in=0] (1.5,7);
\draw[name path=C7--D7,thick,red] (4,5.5) to[out=90, in=-90] (3.5,7.5);
\draw[name path=C8--D8,thick,red] (5,5.5) to[out=90, in=-90] (5.5,7.5);
\draw[name path=C9--D9,thick,red] (5.5,3) to[out=0, in=180] (7.5,2);
\draw[name path=C10--D10,thick,red] (5.5,4) to[out=0, in=180] (7.5,4);
\draw[name path=C11--D11,thick,red] (5.5,5) to[out=0, in=180] (7.5,6);
\draw[name path=C12--D12,thick,red] (2,1.5) to[out=-90, in=90] (1.5,-.5);
\draw[name path=C13--D13,thick,red] (3,1.5) to[out=-90, in=90] (3.5,-.5);
\draw[name path=C14--D14,thick,red] (3.5,2) to[out=0,in=180] (5.5,0);
\draw[name path=C15--D15,thick,red] (4,2.5) to[out=-90, in=90] (6,.5);
\draw[name path=C16--D16,thick,red] (5,2.5) to[out=-90, in=90] (7,.5);
\foreach \x in {1,...,16}{
\path [name intersections={of=A--B and C\x--D\x,by=E\x}];
\filldraw[red] (E\x) circle (.1cm);
}
} 
\,\right\}.
$$
Here, both the internal and external boundaries are given by $\partial \Lambda$, and the $\operatorname{Tube}_\cC(\partial \Lambda)$-action means resolving the $\cC$-morphism from the annulus into the outer-most vertex spaces of $\cH(\Lambda)$.
The object on the boundary is $X^{\otimes \partial \Lambda}$, the tensor product of a copy of $X$ for each site on $\partial \Lambda$.
Thus
\begin{equation}
\label{eq:LargerMETube}
\Tube_\cC(\partial \Lambda)\cong \End_{Z(\cC)}(\Tr(X^{\otimes \partial \Lambda})).
\end{equation}

Applying the gauging unitary takes vectors in the spine coral skein module localized in $\Lambda$ to $\cH(\Lambda)$. 
Combining \eqref{eq:IntermediateSpace2} and \eqref{eq:LargerMETube} with the Yoneda Lemma, we see that conjugating the local operators $\fA(\Lambda)$ which commute with the $\Tube(\cC)$-action on the ungauged side gives the local operators on $\cH(\Lambda)$ which commute with the $\Tube_\cC(\partial\Lambda)$-action.

\subsection{Bounded spread isomorphism of quasi-local operator algebras of observables}
\label{sec:QuasiLocalOperators}

The \emph{quasi-local algebra} of observables $\fA$ is the inductive limit $\rmC^*$-algebra of the local algebras $\fA(\Lambda)$ as the region $\Lambda$ goes to $\infty$, i.e.
$$
\fA := \varinjlim \fA(\Lambda).
$$
We can describe this quasi-local algebra more concretely in terms of inductive limits of certain subalgebras $\fB(\Lambda)\subset B(\cH(\Lambda))$.

For ease of exposition, we now restrict to the case that our bounded contractible patch $\Lambda$ is always a rectangle.
We write $\Lambda^{+1}$ for the rectangle obtained from $\Lambda$ by adding all sites at most distance $\sqrt{2}$ from $\Lambda$.
\begin{equation}
\label{eq:Lambda+1}    
\tikzmath[scale=.5]{
\draw[step=1.0,red,thick] (0.5,0.5) grid (6.5,6.5);
\foreach \i in {1,2,...,6} {
\foreach \j in {1,2,...,6} {
        \filldraw[thick, red] (\i,\j)circle (.1cm);
}}
\draw[thick, rounded corners=5pt] (1.5,1.5) rectangle (5.5,5.5);
\draw[thick, rounded corners=5pt] (2.5,2.5) rectangle (4.5,4.5);
\node at (3.5,3.5) {$\scriptstyle \Lambda$};
\node at (2.1,4.6) {$\scriptstyle \Lambda^{+1}$};
}
\end{equation}

Observe that $\cH(\Lambda)$ decomposes as an orthogonal direct sum into the subspaces with fixed simple labels along the boundary of $\Lambda$.
We say a \emph{boundary condition} for $\cH(\Lambda)$ is an assignment of a simple $c\in \Irr(\cC)$ to each edge
connected to a vertex in $\Lambda$ which is not entirely contained in $\Lambda$, i.e.~the edges which intersect the black line in \eqref{eq:SampleLambda} or \eqref{eq:Lambda+1}.
By an abuse of notation, we denote the set of boundary conditions for $\Lambda$ by $\partial\Lambda$.
We define $\fB(\Lambda)$ to be the local operators on $\cH(\Lambda)$ which preserve each boundary condition; observe that $\fB(\Lambda)$ is isomorphic to a direct sum of full matrix algebras whose summands are labeled by individual boundary conditions:
$$
\fB(\Lambda) \cong \bigoplus_{\vec{c}\in \partial \Lambda} M_{n_{\vec{c}}}(\bbC).
$$

Now consider the two sequences of finite dimensional $\rmC^*$-algebras
\begin{align*}
&\fA(\Lambda)&&\subset&& \fA(\Lambda^{+1})&&\subset&&\fA(\Lambda^{+2})&&\subset&& \cdots
\\
&\fB(\Lambda)&&\subset&& \fB(\Lambda^{+1})&&\subset&&\fB(\Lambda^{+2})&&\subset&& \cdots.
\end{align*}
First, since $\fA(\Lambda)$ is the commutant of the $\Tube_\cC(\partial \Lambda)$-action on $\cH(\Lambda)$, $\fA(\Lambda)$ preserves boundary conditions, so $\fA(\Lambda)\subset \fB(\Lambda)$.
Second, the image of $\fB(\Lambda)$ in $\fB(\Lambda^{+1})$ commutes with the $\Tube_\cC(\partial \Lambda^{+1})$-action on $\cH(\Lambda^{+1})$, and thus lies in $\fA(\Lambda^{+1})$.
Thus the local nets $\fA$ and $\fB$ are actually intertwined
$$
\fA(\Lambda)\subset \fB(\Lambda)\subset \fA(\Lambda^{+1})\subset \fB(\Lambda^{+1})\subset  \cdots,
$$
so the inductive limits are isomorphic.
$$
\fA=\varinjlim \fA(\Lambda) \cong \varinjlim \fB(\Lambda) =: \fB
$$
We thus see we have a bounded spread isomorphism in the sense of \cite{2304.00068} between the two nets of local operator algebras from the braided fusion categorical net $\fA$ formed from $Z(\cC)$ with generator the canonical Lagrangian $A$ to the net of local operators on the Levin-Wen model which preserve local boundary conditions.

\subsection{A modified model to see all excitations}
\label{sec:ModifiedModel}
One slight problem with our previous construction is that our gauged model does not generally see all the sectors of the theory.
A state which violates the operator $B_v$ in the ungauged theory may be used as a cyclic vector to generate a $\Tube(\cC)$-module.
In the gauged theory, such a state violates the $B_p$ term in the Hamiltonian induced by the gauging map from the ungauged theory.
By virtue of this state being in the image of the gauging map, it obeys the gauge constraint.
Furthermore, since fluxes are excitations which violate the gauge constraint, we are therefore well-justified in calling this excitation a charge.
However, we know that not all anyons in $Z(\cC)$ are charges.\footnote{In the Levin-Wen model, a \emph{charge} is an anyon $z\in\Irr(Z(\cC))$ corresponding to a violation of a $B_p$ term, a \emph{flux} is an anyon corresponding to a violation of a single edge term $A_\ell$, and a \emph{dyon} is an anyon corresponding to a violation of both edge and plaquette terms.
The operator $A_\ell$ may be interpreted as a the local gauge constraint.
Mathematically, charges are exactly the anyons which admit a non-zero morphism to $A=\Tr(1_\cC)$.} 
This is manifested mathematically by the failure of the canonical Lagrangian $A\in Z(\cC)$ to always be a tensor generator. 
Physically, we see that the full symmetry object $\Tube(\cC)$ does not naturally appear as local operators in the net of gauge invariant operators. 
Instead, the `corner' $p_{1_\cC}\Tube(\cC)p_{1_\cC}$, which is isomorphic to the fusion algebra, naturally appears inside $\End_{Z(\cC)}(A)$ acting on-site. 
This issue is an artifact of restricting to the charge sector of the initial Hilbert space which was necessary to obtain the version of the Levin-Wen model from \cite{MR3204497,2305.14068} from gauging.

However, we can replace the models on both sides of the gauging unitary to see flux and dyon excitations after gauging, and not just charges.
We replace the local on-site Hilbert spaces $\cK_v$ and $\cH_v$ with
\begin{align*}
\cK'_v
&=
\tikzmath{
\draw[thick] (0,0) -- (0,1);
\draw[thick] (1,0) -- (1,1);
\draw[thick, red] (.5,-.2) -- (.5,.8);
\halfDottedEllipse{(0,0)}{.5}{.2}
\halfDottedEllipse[thick, red]{(0,.5)}{.5}{.2}
\filldraw[thick, fill=white] (.5,1) ellipse (.5 and .2);
\fill[red] (.5,.3) circle (.05cm);
}
=
\Tube(\cC)
\\
\cH_v'
&=
\tikzmath{
\draw[blue!30] (.1,.9) -- (.9,.1);
\draw[blue!30,->] (.1,.8) -- (.25,.95);
\draw[blue!30,->] (.8,.1) -- (.95,.25);
\draw[dotted, rounded corners=5pt] (0,0) rectangle (1,1);
\fill[red] (.5,.5) circle (.05cm);
\draw[thick, red] (.5,0) -- (.5,1);
\draw[thick, red] (0,.5) -- (1,.5);
\draw[thick, red] (.5,.5) -- (.3,1);
}
=
\cC(X^{\otimes 2}\to X^{\otimes 3})
\end{align*}
respectively.
The plaquette operators on both sides are now adapted to project the extra copy of $X$ down to $1_\cC$ before applying the usual $B_p$ operators.
Denoting this projection $X\to 1_\cC$ as a univalent vertex, these operators can be represented graphically by
$$
\tikzmath{
\draw[thick] (0,0) -- (0,1);
\draw[thick] (1,0) -- (1,1);
\halfDottedEllipse{(0,0)}{.5}{.2}
\draw[thick, dotted] (.5,1) ellipse (.5 and .2);
\draw[thick] (0,1) arc (180:0:.5 and .4);
\draw[thick] (0,2) arc (-180:0:.5 and .4);
\filldraw[thick, fill=white] (.5,2) ellipse (.5 and .2);
\fill[red] (.5,.5) circle (.05cm);
\draw[thick, red] (.5,.5) -- (.5,-.2);
}
\qquad\text{and}\qquad
\frac{1}{D_\cC}
\sum_{c \in \Irr(\cC)} 
d_c\,
\tikzmath{
\foreach \x in{0,1}{
\foreach \y in{0,1}{
\fill[red] (\x,\y) circle (.05cm);
\draw[thick, red] ($(\x,\y)-(.5,0)$) -- ($(\x,\y)+(.5,0)$);
\draw[thick, red] ($(\x,\y)-(0,.5)$) -- ($(\x,\y)+(0,.5)$);
\draw[thick, red] (\x,\y) -- ($(\x,\y)+(-.1,.2)$);
}}
\fill[red] ($(1,0)+(-.1,.2)$) circle (.05cm);
\draw[thick, blue, rounded corners=5pt] (.1,.2) rectangle (.8,.9);
\node[blue] at (.2,.5) {$\scriptstyle c$};
}
$$
respectively.
The gauging unitary is similar to before, but the local operator algebras of observables are now given by
$$
\fA'(\Lambda)= \End_{Z(\cC)}(\Tr(X)^{\otimes \#p\in\Lambda}).
$$
Since $Z(\cC)(z\to \Tr(X))\neq 0$ for all $z\in\Irr(Z(\cC))$, we now can get every anyon type as violations of the $B_p$ operators acting on $\im(P_A)$ including the fluxes and dyons, and not just charges.

Observe that $\Tube(\cC)\cong\End_{Z(\cC)}(\Tr(X))$ now acts locally on-site after applying the gauging unitary.
Indeed, $\cK_v$ carries two commuting $\Tube(\cC)$-actions, and we only used one of these to form the spine coral skein module.
So the local on-site operators are exactly the operators from the other action.
This on-site $\Tube(\cC)$-action gives a local gauge symmetry action on the gauged Hilbert space.

\section{Group categories}
\label{sec:GroupExample}

We now work through our gauging map for pointed UFCs, i.e., those of the form $\Hilb(G,\omega)$ where $\omega\in Z^3(G,U(1))$.

\subsection{Trivial cocycles}

In the standard story of gauging a $G$-SPT ($\cC=\Hilb(G)$) to produce the Quantum Double, one may start with an SPT with a trivial cocycle. 
In this model, each site has a Hilbert space given by $\mathbb{C}[G]$. The Hamiltonian is given by
$$
H=-\sum\limits_{v\in\Gamma}B_v
$$
where
$$
B_v=\dfrac{1}{|G|}\sum\limits_{g\in G}L_v^g
$$
and $L_v^g$ is the left action of $g\in G$ on the site $v$. The symmetry action $U^g$ for $g\in G$ is given by the right action of $g$ on each site.

When $\cC=\Hilb(G)$, $gg^{-1}=e$ implies
$$
\cK_v
=
\tikzmath{
\draw[thick] (0,0) -- (0,1);
\draw[thick] (1,0) -- (1,1);
\halfDottedEllipse{(0,0)}{.5}{.2}
\halfDottedEllipse[thick, red]{(0,.5)}{.5}{.2}
\filldraw[thick, fill=white] (.5,1) ellipse (.5 and .2);
\fill[red] (.5,.3) circle (.05cm);
\draw[thick, red] (.5,.3) -- (.5,-.2);
}
=
\operatorname{span}
\set{
\tikzmath{
\draw[thick] (0,0) -- (0,1);
\draw[thick] (1,0) -- (1,1);
\halfDottedEllipse{(0,0)}{.5}{.2}
\halfDottedEllipse[thick, blue]{(0,.5)}{.5}{.2}
\node[blue] at (.5,.45) {$\scriptstyle g$};
\filldraw[thick, fill=white] (.5,1) ellipse (.5 and .2);
}~
}{~g\in G}
=
\bbC[G],
$$
which is exactly
the fusion/group algebra.
Moreover, the tube algebra action factors through the forgetful image of the fusion algebra $\bbC[G]$.
The image of the spine coral is what you get using empty coral, which is exactly the tensor product in $\Hilb$, and we get the diagonal $\bbC[G]$-action.
This is because of the recabling relation:
$$
\tikzmath{
\draw[thick, blue, mid>] (-.3,-.5) --node[left]{$\scriptstyle g$} (-.3,.5);
\draw[thick, blue, mid<] (.3,-.5) -- (.3,.5);
}
=
\tikzmath{
\draw[thick, blue, mid>] (-.3,-.5) node[left, yshift=.2cm]{$\scriptstyle g$} arc(180:0:.3cm);
\draw[thick, blue, mid<] (-.3,.5) node[left, yshift=-.2cm]{$\scriptstyle g$} arc(-180:0:.3cm);
}
$$
which proves that we can achieve the tensor product of $\bbC[G]$ representations using the pair of pants.
$$ 
\tikzmath[yscale=-1]{
\draw[thick] (0,0) -- (0,-1);
\draw[thick] (.6,0) -- (.6,-1);
\draw[thick] (1.4,0) -- (1.4,-1);
\draw[thick] (2,0) -- (2,-1);
\draw[thick] (.3,-1) ellipse (.3 and .1);
\draw[thick] (1.7,-1) ellipse (.3 and .1);
\colorHalfDottedEllipseTwo{(0,-.5)}{.3}{.1}{blue}
\node[blue] at (.3,-.25) {$\scriptstyle g$};
\colorHalfDottedEllipseTwo{(1.4,-.5)}{.3}{.1}{blue}
\node[blue] at (1.7,-.25) {$\scriptstyle g$};
\pairOfPantsTwo{(0,0)};
}
\quad
=
\tikzmath[yscale=-1]{
\halfDottedEllipseTwo{(.7,2.5)}{.3}{.1}
\draw[thick] (.7,1.5) -- (.7,2.5);
\draw[thick] (1.3,1.5) -- (1.3,2.5);
\colorHalfDottedEllipseTwo{(.7,2)}{.3}{.1}{blue}
\node[blue] at (1,2.25) {$\scriptstyle g$};
\topPairOfPantsTwo{(0,0)}
}
$$
Thus taking fixed points is exactly the usual SPT space.
The gauging map is the usual well-known one; the on-site vector in $\cH_v$ obtained by applying the gauging map to a simple tensor is given as follows. 
$$
\tikzmath{
\draw[dashed] (-.5,.75) -- (2.5,.75);
\draw[dashed] (1,-.5) -- (1,2);
\foreach \x in {0,2}{
\foreach \y in {0,1.5}{
\draw[thick] (\x,\y) circle (.3cm);
\draw[thick,blue] (\x,\y) circle (.5cm);
}}
\node[blue] at (-.65,0) {$\scriptstyle g$};
\node[blue] at (.75,0) {$\scriptstyle g^{-1}$};
\node[blue] at (1.35,0) {$\scriptstyle h$};
\node[blue] at (2.75,0) {$\scriptstyle h^{-1}$};
\node[blue] at (-.65,1.5) {$\scriptstyle \ell$};
\node[blue] at (.75,1.5) {$\scriptstyle \ell^{-1}$};
\node[blue] at (1.35,1.5) {$\scriptstyle k$};
\node[blue] at (2.75,1.5) {$\scriptstyle k^{-1}$};
}
\rightsquigarrow
\tikzmath{
\draw[thick, blue] (0,1) node[left]{$\scriptstyle \ell^{-1}g$} -- (2,1) node[right]{$\scriptstyle k^{-1}h$};
\draw[thick, blue] (1,0) node[below]{$\scriptstyle g^{-1}h$} -- (1,2)node[above]{$\scriptstyle \ell^{-1}k$};
}
$$
Above, the dashed line corresponds to the Levin-Wen Hilbert space, and the squiggly arrow represents resolving the string diagram onto the Levin-Wen grid in the usual way, where the string states are given by $G$-valued domain walls from the SPT states.
We give full details on how to interpret the above diagram in the next subsection, where we no longer assume the cocycle is trivial.

\subsection{Twisted Gauging}

Using our gauging map, the case of $\cC=\Hilb(G,\omega)$ with $\omega\in Z^3(G,U(1))$  is very similar to $\cC=\Hilb(G)$. 
Again, the fact that $gg^{-1}=e$ implies that $\cK_v=\bbC[G]$ and that the action of the tube algebra factors through the forgetful image of the fusion algebra $\bbC[G]$.
We again get that the tensor product of the $\cK_v$ is the tensor product in $\Hilb$ with the diagonal $\bbC[G]$-action due to the recabling relation (see \eqref{eq:RecablingWithCocycle} below), even though there is no fiber functor from $\Hilb(G,\omega)$ to $\Hilb$ when $\omega$ is non-trivial.
Indeed, $\Hilb(G,\omega)$ is $G$-graded,
and thus $Z(\Hilb(G,\omega))$ contains a canonical copy of $\Rep(G)$ 
whose image under the forgetful functor lies over the unit object \cite{MR2587410}, and the canonical Lagrangian algebra $A$ corresponding to $\Hilb(G,\omega)$ is again the function algebra in this copy of $\Rep(G)$.
The big difference, however, is that the cocycle $\omega$ serves to twist the gauging map and, therefore, the gauge symmetry of the string-net model.

We now review the diagrammatic presentation for $\Hilb(G,\omega)$.
There is a strand for each group element and a distinguished unitary trivalent vertex from $\mu_{g,h};gh\to g\otimes h$ for each $g,h\in G$ satisfying various relations.
We may assume $\mu_{g,e}=\mu_{e,g}=\id_g$ under identifying $g\otimes e = g = e\otimes g$ using unitors.
We represent $\mu_{g,h}$ and $\mu_{g,h}^\dag$ by
$$
\tikzmath{
\draw[thick, blue] (-.3,.3) node[above]{$\scriptstyle g$} -- (0,0) -- (.3,.3) node[above]{$\scriptstyle h$};
\draw[thick, blue] (0,0) -- (0,-.3) node[below]{$\scriptstyle gh$};
\filldraw[blue] (0,0) node[left]{$\scriptstyle \mu_{g,h}$} circle (.05cm);
}
\qquad\qquad
\tikzmath{
\draw[thick, blue] (-.3,-.3) node[below]{$\scriptstyle g$} -- (0,0) -- (.3,-.3) node[below]{$\scriptstyle h$};
\draw[thick, blue] (0,0) -- (0,.3) node[above]{$\scriptstyle gh$};
\filldraw[blue] (0,0) node[left,xshift=-.1cm]{$\scriptstyle \mu_{g,h}^\dag$} circle (.05cm);
}\,.
$$
The cocycle appears in the associativity conditions 
\begin{align*}
\tikzmath{
\draw[thick, blue] (.3,.3) node[above]{$\scriptstyle k$} -- (0,0) -- (-.3,.3) node[above]{$\scriptstyle h$};
\draw[thick, blue] (0,0) -- (-.3,-.3) -- (-.9,.3) node[above]{$\scriptstyle g$};
\draw[thick, blue] (-.3,-.3) -- (-.3,-.6) node[below]{$\scriptstyle ghk$};
\filldraw[blue] (0,0) node[right]{$\scriptstyle \mu_{h,k}$} circle (.05cm);
\filldraw[blue] (-.3,-.3) node[right]{$\scriptstyle \mu_{g,hk}$} circle (.05cm);
}
&=
\omega_{g,h,k}
\tikzmath{
\draw[thick, blue] (-.3,.3) node[above]{$\scriptstyle g$} -- (0,0) -- (.3,.3) node[above]{$\scriptstyle h$};
\draw[thick, blue] (0,0) -- (.3,-.3) -- (.9,.3) node[above]{$\scriptstyle k$};
\draw[thick, blue] (.3,-.3) -- (.3,-.6) node[below]{$\scriptstyle ghk$};
\filldraw[blue] (0,0) node[left]{$\scriptstyle \mu_{g,h}$} circle (.05cm);
\filldraw[blue] (.3,-.3) node[left]{$\scriptstyle \mu_{gh,k}$} circle (.05cm);
}
\\
\tikzmath[yscale=-1]{
\draw[thick, blue] (.3,.3) node[below]{$\scriptstyle k$} -- (0,0) -- (-.3,.3) node[below]{$\scriptstyle h$};
\draw[thick, blue] (0,0) -- (-.3,-.3) -- (-.9,.3) node[below]{$\scriptstyle g$};
\draw[thick, blue] (-.3,-.3) -- (-.3,-.6) node[above]{$\scriptstyle ghk$};
\filldraw[blue] (0,0) node[right]{$\scriptstyle \mu_{h,k}^
\dag$} circle (.05cm);
\filldraw[blue] (-.3,-.3) node[right]{$\scriptstyle \mu_{g,hk}^
\dag$} circle (.05cm);
}
&=
\omega_{g,h,k}^{-1}
\tikzmath[yscale=-1]{
\draw[thick, blue] (-.3,.3) node[below]{$\scriptstyle g$} -- (0,0) -- (.3,.3) node[below]{$\scriptstyle h$};
\draw[thick, blue] (0,0) -- (.3,-.3) -- (.9,.3) node[below]{$\scriptstyle k$};
\draw[thick, blue] (.3,-.3) -- (.3,-.6) node[above]{$\scriptstyle ghk$};
\filldraw[blue] (0,0) node[left]{$\scriptstyle \mu_{g,h}^
\dag$} circle (.05cm);
\filldraw[blue] (.3,-.3) node[left]{$\scriptstyle \mu_{gh,k}^
\dag$} circle (.05cm);
}.
\end{align*}
This immediately implies the following triangle evaluation relations.
\begin{align*}
\tikzmath{
\draw[thick, blue] (0,0) -- (0,1.2);
\draw[thick, blue] (0,.3) -- (.3,.6) -- (0,.9);
\draw[thick, blue] (.3,.6) -- (.9,1.2);
\filldraw[blue] (0,.3) node[left]{$\scriptstyle \mu_{g,hk}$} circle (.05cm);
\filldraw[blue] (.3,.6) node[right]{$\scriptstyle \mu_{h,k}$} circle (.05cm);
\filldraw[blue] (0,.9) node[left]{$\scriptstyle \mu^\dag_{g,h}$} circle (.05cm);
}
&=
\omega_{g,h,k}
\tikzmath{
\draw[thick, blue] (-.3,.3) node[above]{$\scriptstyle gh$} -- (0,0) -- (.3,.3) node[above]{$\scriptstyle k$};
\draw[thick, blue] (0,0) -- (0,-.3) node[below]{$\scriptstyle ghk$};
\filldraw[blue] (0,0) node[left]{$\scriptstyle \mu_{gh,k}$} circle (.05cm);
}
\\
\tikzmath{
\draw[thick, blue] (0,0) -- (0,1.2);
\draw[thick, blue] (0,.3) -- (-.3,.6) -- (0,.9);
\draw[thick, blue] (-.3,.6) -- (-.9,1.2);
\filldraw[blue] (0,.3) node[right]{$\scriptstyle \mu_{gh,k}$} circle (.05cm);
\filldraw[blue] (-.3,.6) node[left]{$\scriptstyle \mu_{g,h}$} circle (.05cm);
\filldraw[blue] (0,.9) node[right]{$\scriptstyle \mu^\dag_{h,k}$} circle (.05cm);
}
&=
\omega_{g,h,k}^{-1}
\tikzmath{
\draw[thick, blue] (-.3,.3) node[above]{$\scriptstyle g$} -- (0,0) -- (.3,.3) node[above]{$\scriptstyle hk$};
\draw[thick, blue] (0,0) -- (0,-.3) node[below]{$\scriptstyle ghk$};
\filldraw[blue] (0,0) node[left]{$\scriptstyle \mu_{g,hk}$} circle (.05cm);
}
\end{align*}

Without loss of generality, we may assume $\omega$ is normalized so that whenever one of its arguments is the identity $e$, the value is $1$.
This immediately implies $\omega_{g,g^{-1},g}=\omega_{g^{-1},g,g^{-1}}^{-1}$ for all $g\in G$ by expanding $d\omega_{g,g^{-1},g,g^{-1}}=1$, so the duality maps can be chosen to be
\begin{align*}
\tikzmath{
\draw[thick,blue] (0,0) node[below]{$\scriptstyle g^{-1}$} arc (180:0:.3cm) node[below]{$\scriptstyle g$};
}
:&=
\tikzmath{
\draw[thick, blue] (-.3,-.3) node[below]{$\scriptstyle g^{-1}$} -- (0,0) -- (.3,-.3) node[below]{$\scriptstyle g$};
\draw[thick, blue, dotted] (0,0) -- (0,.3) node[above]{$\scriptstyle e$};
\filldraw[blue] (0,0) node[left,xshift=-.1cm]{$\scriptstyle \mu_{g^{-1},g}^\dag$} circle (.05cm);
}
\\
\tikzmath{
\draw[thick,blue] (0,0) node[above]{$\scriptstyle g$} arc (-180:0:.3cm) node[above]{$\scriptstyle g^{-1}$};
}
:&=
\omega_{g,g^{-1},g}
\tikzmath{
\draw[thick, blue] (-.3,.3) node[above]{$\scriptstyle g^{-1}$} -- (0,0) -- (.3,.3) node[above]{$\scriptstyle g$};
\draw[thick, blue, dotted] (0,0) -- (0,-.3) node[below]{$\scriptstyle e$};
\filldraw[blue] (0,0) node[right]{$\scriptstyle \mu_{g,g^{-1}}$} circle (.05cm);
}\,.
\end{align*}
The unique unitary spherical structure is given by $\varphi_g:= \omega_{g^{-1},g,g^{-1}}$ so that $d_g=1$:
$$
\tikzmath{
\draw[thick, blue] (0,.3) arc (180:0:.3cm) --node[right]{$\scriptstyle g^{-1}$} (.6,-.3) arc (0:-180:.3cm);
\roundNbox{blue}{(0,0)}{.3}{0}{0}{\textcolor{blue}{$\varphi_g$}}
}
=
\omega_{g,g^{-1},g}\cdot
\omega_{g^{-1},g,g^{-1}}\cdot
\tikzmath{
\draw[thick, blue] (0,0) node[left]{$\scriptstyle g$} arc (180:0:.3cm) node[right]{$\scriptstyle g^{-1}$} arc (0:-180:.3cm);
\filldraw[blue] (.3,.3) circle (.05cm);
\filldraw[blue] (.3,-.3) circle (.05cm);
\draw[thick, blue, dotted] (.3,.3) -- (.3,.6) node[above]{$\scriptstyle e$};
\draw[thick, blue, dotted] (.3,-.3) -- (.3,-.6) node[below]{$\scriptstyle e$};
}
=
\id_e.
$$
We immediately see we have the usual recabling relation, where the same two scalars that appear in the line above again cancel:
\begin{equation}
\label{eq:RecablingWithCocycle}
\tikzmath{
\draw[thick, blue] (-.3,-.5) node[below]{$\scriptstyle g$} -- (-.3,.5) node[above]{$\scriptstyle g$};
\draw[thick, blue] (.3,-.5)node[below]{$\scriptstyle g^{-1}$} -- (.3,.5) node[above]{$\scriptstyle g^{-1}$};
}
=
\tikzmath{
\draw[thick, blue] (0,.3) node[below]{$\scriptstyle g$} arc(180:0:.3) node[below]{$\scriptstyle g^{-1}$};
\draw[thick, blue] (0,1.2) node[above]{$\scriptstyle g$} arc(-180:0:.3) node[above]{$\scriptstyle g^{-1}$};
\draw[thick, blue,dotted] (.3,.6) -- (.3,.9);
\filldraw[blue] (.3,.6) circle (.05cm);
\filldraw[blue] (.3,.9) circle (.05cm);
}
=\,\,
\tikzmath{
\draw[thick, blue] (0,-.3) -- (0,-.6) node[below]{$\scriptstyle g$};
\draw[thick, blue] (0,.3) arc(180:0:.3) -- (.6,-.6) node[below]{$\scriptstyle g^{-1}$};
\draw[thick, blue] (0,1.2) node[above]{$\scriptstyle g$} arc(-180:0:.3) node[above]{$\scriptstyle g^{-1}$};
\roundNbox{blue}{(0,0)}{.3}{0}{0}{\textcolor{blue}{$\varphi_g$}}
}\,.
\end{equation}
Another way to see both $d_g=1$ and \eqref{eq:RecablingWithCocycle} is that
$$
\left(
\tikzmath{
\draw[thick,blue] (0,0) node[above]{$\scriptstyle g$} arc (-180:0:.3cm) node[above]{$\scriptstyle g^{-1}$};
}
\right)^\dag
=
\tikzmath{
\draw[thick, blue] (0,-.3) -- (0,-.6) node[below]{$\scriptstyle g$};
\draw[thick, blue] (0,.3) arc(180:0:.3) -- (.6,-.6) node[below]{$\scriptstyle g^{-1}$};
\roundNbox{blue}{(0,0)}{.3}{0}{0}{\textcolor{blue}{$\varphi_g$}}
}\,.
$$
It is important to remember that these trivalent vertices are not preserved under rotation, i.e., in general, rotating an output strand of $\mu_{g,h}$ downward will introduce a cocycle value times the corresponding $\mu^\dag$.

When our cocycle is non-trivial, our gauging map differs 
on simple tensors in the canonical basis to the on-site Levin-Wen Hilbert space $\cH_v$ by a scalar $\lambda_{g,h,k,\ell}$ which is a product of cocycle values depending on the four group elements.
In this sense, we say that the cocycle twists the gauging map.
$$
\tikzmath{
\draw[dashed] (-.5,.75) -- (2.5,.75);
\draw[dashed] (1,-.5) -- (1,2);
\foreach \x in {0,2}{
\foreach \y in {0,1.5}{
\draw[thick] (\x,\y) circle (.3cm);
\draw[thick,blue] (\x,\y) circle (.5cm);
}}
\node[blue] at (-.65,0) {$\scriptstyle g$};
\node[blue] at (.75,0) {$\scriptstyle g^{-1}$};
\node[blue] at (1.35,0) {$\scriptstyle h$};
\node[blue] at (2.75,0) {$\scriptstyle h^{-1}$};
\node[blue] at (-.65,1.5) {$\scriptstyle \ell$};
\node[blue] at (.75,1.5) {$\scriptstyle \ell^{-1}$};
\node[blue] at (1.35,1.5) {$\scriptstyle k$};
\node[blue] at (2.75,1.5) {$\scriptstyle k^{-1}$};
}
\rightsquigarrow
\lambda_{g,h,k,\ell}\cdot
\tikzmath{
\draw[thick, blue] (0,.5) node[left]{$\scriptstyle \ell^{-1}g$} -- (1,.5) node[right]{$\scriptstyle k^{-1}h$};
\draw[thick, blue] (.5,0) node[below]{$\scriptstyle g^{-1}h$} -- (.5,1) node[above]{$\scriptstyle \ell^{-1}k$};
}
$$
Above, the 4-valent vertex is further resolved as
$$
\tikzmath{
\draw[thick, blue] (-.5,0) node[left]{$\scriptstyle \ell^{-1}g$} -- (.5,0) node[right]{$\scriptstyle k^{-1}h$};
\draw[thick, blue] (0,-.5) node[below]{$\scriptstyle g^{-1}h$} -- (0,.5) node[above]{$\scriptstyle \ell^{-1}k$};
}
=
\tikzmath{
\draw[thick, blue] (-.5,0) node[left]{$\scriptstyle \ell^{-1}g$} -- (0,0) node[right]{$\scriptstyle \mu^\dag_{\ell^{-1}g,g^{-1}h}$} -- (0,-.5) node[below]{$\scriptstyle g^{-1}h$};
\draw[thick, blue] (0,0) --node[left, yshift=.1cm]{$\scriptstyle \ell^{-1}h$} (1,1);
\draw[thick, blue] (1,1.5) node[above]{$\scriptstyle \ell^{-1}k$} -- (1,1) node[left]{$\scriptstyle \mu_{\ell^{-1}k,k^{-1}h}$} -- (1.5,1) node[right]{$\scriptstyle k^{-1}h$};
\filldraw[blue] (0,0) circle (.05cm);
\filldraw[blue] (1,1) circle (.05cm);
}\,.
$$
The scalar $\lambda_{g,h,k,\ell}$ is defined by 
$$
\tikzmath{
\filldraw[blue] (0,.4) circle (.05cm);
\filldraw[blue] (-1,1) circle (.05cm);
\filldraw[blue] (1,1) circle (.05cm);
\filldraw[blue] (0,-.4) circle (.05cm);
\filldraw[blue] (-1,-1) circle (.05cm);
\filldraw[blue] (1,-1) circle (.05cm);
\draw[thick, blue] (-1,-1.3) node[below]{$\scriptstyle \ell^{-1}g$} --node[left]{$\scriptstyle \ell^{-1}$} (-1,1.3) node[above]{$\scriptstyle \ell^{-1}k$};
\draw[thick, blue] (1,-1.3) node[below]{$\scriptstyle g^{-1}h$} --node[right]{$\scriptstyle h$} (1,1.3) node[above]{$\scriptstyle k^{-1}h$};
\draw[thick, blue] (-1,1) --node[above]{$\scriptstyle k$} (0,.4) --node[above]{$\scriptstyle k^{-1}$} (1,1);
\draw[thick, blue] (-1,-1) --node[below]{$\scriptstyle g$} (0,-.4) --node[below, xshift=-.1cm]{$\scriptstyle g^{-1}$} (1,-1);
\draw[thick, blue, dotted] (0,-.4) -- (0,.4);
}
=
\lambda_{g,h,k,\ell}\cdot
\tikzmath{
\draw[thick, blue] (-.3,-.3) node[below,xshift=-.2cm]{$\scriptstyle \ell^{-1}g$} -- (0,0) -- (.3,-.3) node[below,xshift=.2cm]{$\scriptstyle g^{-1}h$};
\draw[thick, blue] (-.3,.9) node[above,xshift=-.2cm]{$\scriptstyle \ell^{-1}k$} -- (0,.6) -- (.3,.9) node[above,xshift=.2cm]{$\scriptstyle k^{-1}h$};
\draw[thick, blue] (0,0) --node[left]{$\scriptstyle \ell^{-1}h$} (0,.6);
\filldraw[blue] (0,0) node[right]{$\scriptstyle \mu_{\ell^{-1}g,g^{-1}h}^\dag$} circle (.05cm);
\filldraw[blue] (0,.6) node[right]{$\scriptstyle \mu_{\ell^{-1}k,k^{-1}h}$} circle (.05cm);
}
$$
and can be shown to be equal to
$$
\omega_{\ell^{-1},k,k^{-1}}
\cdot
\omega^{-1}_{\ell^{-1},g,g^{-1}}
\cdot
\omega_{\ell^{-1}g,g^{-1},h}
\cdot
\omega^{-1}_{\ell^{-1}k,k^{-1},h}
$$
by applying the fusion and triangle evaluation relations.

We can now resolve the following simple tensor in the spine coral skein module 
$$
\tikzmath{
\coordinate (a) at (.8,-3);
\coordinate (b11) at (-.5,0);
\coordinate (b21) at (1.5,0);
\coordinate (b12) at (-.25,1.5);
\coordinate (b22) at (1.75,1.5);
\halfDottedEllipse{(a)}{.5}{.2}
\draw[thick] (b12) -- ($ (b12) + (0,-.5) $) .. controls ++(270:.4cm) and ++(90:.4cm) .. ($ (b11) + (.6,0) $) ;
\draw[thick] ($ (b11) + (1,0) $) -- ($ (b11) + (1,-.5) $) .. controls ++(270:.3cm) and ++(110:.3cm) .. ($ (b11) + (1.3,-1.2) $) ;
\draw[thick] ($ (b21) + (1,0) $) -- ($ (b21) + (1,-.5) $) .. controls ++(270:.3cm) and ++(70:.3cm) .. ($ (b21) + (.7,-1.2) $) ;
\draw[thick] (b21) -- ($ (b21) + (0,-.5) $) .. controls ++(270:.3cm) and ++(70:.3cm) .. ($ (b21) + (-.3,-1.2) $) ;
\draw[thick] (a) .. controls ++(90:.8cm) and ++(270:1.2cm) .. ($ (b11) + (0,-.5) $) -- (b11);
\draw[thick] ($ (a) + (1,0) $) .. controls ++(90:.8cm) and ++(270:1.2cm) .. (2.8,-1) -- ($ (b22) + (1,0) $);
\draw[thick] ($ (b12) + (1,0) $) -- ($ (b12) + (1,-.5) $) arc (-180:0:.5) -- (b22);
\draw[thick] (a) arc (-180:0:.5cm);
\foreach \x in {0,2}{
\foreach \y in {0,1.5}{
\halfDottedEllipse[thick, blue]{(\x+\y*.25/1.5-.5,\y-.5)}{.5}{.2}
}}
\foreach \x in {0,2}{
\foreach \y in {0,1.5}{
\filldraw[thick, fill=white] (\x+\y*.25/1.5,\y) ellipse (.5 and .2);
}}
\node[blue] at (0,-.9) {$\scriptstyle g$};
\node[blue] at (2,-.9) {$\scriptstyle h$};
\node[blue] at (.25,.6) {$\scriptstyle \ell$};
\node[blue] at (2.25,.6) {$\scriptstyle k$};
}
$$
as the following vector in the Levin-Wen Hilbert space:
\begin{align*}
\textcolor{blue}{\lambda_{g,h,k,\ell}}
\cdot
\textcolor{red}{\lambda_{e,e,h,g}}
\cdot
\textcolor{orange}{\lambda_{h,e,e,k}}
\cdot
\textcolor{violet}{\lambda_{\ell,k,e,e}}
\cdot
\textcolor{black}{\lambda_{e,g,\ell,e}}
\cdot
\\
\tikzmath{
\draw[thick, blue] (0,0) rectangle (4,4);
\draw[thick, blue] (2,0) -- (2,4);
\draw[thick, blue] (0,2) -- (4,2);
\node[blue] at (1,.2) {$\scriptstyle g^{-1}$};
\node[blue] at (3,.2) {$\scriptstyle h^{-1}$};
\node[blue] at (.2,1) {$\scriptstyle g$};
\node[blue] at (2.4,1) {$\scriptstyle g^{-1}h$};
\node[blue] at (4.3,1) {$\scriptstyle h^{-1}$};
\node[blue] at (1,2.2) {$\scriptstyle \ell^{-1}g$};
\node[blue] at (3,2.2) {$\scriptstyle k^{-1}h$};
\node[blue] at (.2,3) {$\scriptstyle \ell$};
\node[blue] at (2.4,3) {$\scriptstyle \ell^{-1}k$};
\node[blue] at (4.3,3) {$\scriptstyle k^{-1}$};
\node[blue] at (1,4.2) {$\scriptstyle \ell$};
\node[blue] at (3,4.2) {$\scriptstyle k$};
\node[blue] at (-.2,-.2) {$\scriptstyle \mu_{g,g^{-1}}$};
\node[blue] at (-.5,2) {$\scriptstyle \textcolor{black}{\mu_{\ell,\ell^{-1}g}}$};
\node[blue] at (-.2,4.2) {$\scriptstyle \id_\ell$};
\node[blue] at (2,-.2) {$\scriptstyle \textcolor{red}{\mu_{g^{-1}h,h^{-1}}}$};
\node[blue] at (2,4.3) {$\scriptstyle \textcolor{violet}{\mu^\dag_{\ell,\ell^{-1}k}}$};
\node[blue] at (4.2,-.2) {$\scriptstyle \id_{h^{-1}}$};
\node[blue] at (4.7,2) {$\scriptstyle \textcolor{orange}{\mu^\dag_{k^{-1}h,h^{-1}}}$};
\node[blue] at (4.7,4.2) {$\scriptstyle \mu^\dag_{k,k^{-1}}$};
}.
\end{align*}
The middle 4-valent vertex contributes $\lambda_{g,h,k,\ell}$ giving 4 cocycle values, and the four trivalent vertices each contribute a degenerate $\lambda$ constant giving one cocycle value apiece.
The 4 corners also each contribute a degenerate $\lambda$ constant equal to 1.

\section{The Fibonacci category}
\label{sec:FibExample}
In the interest of having a non group-theoretical example, let $\cC=\mathsf{Fib}$, the Fibonacci category, whose simple objects are $1$ and $\tau$, subject to the fusion rules $\tau\otimes\tau=1\oplus\tau$.  The object $F(A)$ is given by
\[F(A)=F\big(\Tr(1)\big)=1\otimes1\,\oplus\,\tau\otimes\tau=2\cdot1\,\oplus\,\tau\,.\]
This implies that
$$
\tikzmath{
\draw[thick] (0,0) -- (0,1);
\draw[thick] (1,0) -- (1,1);
\halfDottedEllipse{(0,0)}{.5}{.2}
\halfDottedEllipse[thick, red]{(0,.5)}{.5}{.2}
\filldraw[thick, fill=white] (.5,1) ellipse (.5 and .2);
\fill[red] (.5,.3) circle (.05cm);
\draw[thick, red] (.5,.3) -- (.5,-.2);
}
=
\operatorname{span}
\left\{
\tikzmath{
\draw[thick] (0,0) -- (0,1);
\draw[thick] (1,0) -- (1,1);
\halfDottedEllipse{(0,0)}{.5}{.2}
\draw[thick, blue, dotted] (0.5,0.5) ellipse (.5 and .2);
\node[blue] at (.5,.45) {$\scriptstyle 1$};
\filldraw[thick, fill=white] (.5,1) ellipse (.5 and .2);
}\;,~
\tikzmath{
\draw[thick] (0,0) -- (0,1);
\draw[thick] (1,0) -- (1,1);
\halfDottedEllipse{(0,0)}{.5}{.2}
\halfDottedEllipse[thick, blue]{(0,.5)}{.5}{.2}
\node[blue] at (.5,.45) {$\scriptstyle \tau$};
\filldraw[thick, fill=white] (.5,1) ellipse (.5 and .2);
}\;,~
\tikzmath{
\draw[thick] (0,0) -- (0,1);
\draw[thick] (1,0) -- (1,1);
\halfDottedEllipse{(0,0)}{.5}{.2}
\halfDottedEllipse[thick, blue]{(0,.5)}{.5}{.2}
\node[blue] at (.3,.45) {$\scriptstyle \tau$};
\node[blue] at (.65,.05) {$\scriptstyle \tau$};
\filldraw[thick, fill=white] (.5,1) ellipse (.5 and .2);
\fill[blue] (.5,.3) circle (.05cm);
\draw[thick, blue] (.5,.3) -- (.5,-.2);
}~
\right\}\,.
$$

We now include a specific example of our gauging map.
Recall that $\mathsf{Fib}$ has a diagrammatic presentation where the $\tau$-strand satisfies the skein relations
\begin{align}
\label{eq:FibSkein1}
\tikzmath{
\draw[thick, blue] (-.3,-.6) -- (-.3,.6);
\draw[thick, blue] (.3,-.6) -- (.3,.6);
}\,
&=
\frac{1}{\phi}
\tikzmath{
\draw[thick, blue] (-.3,.6) arc (-180:0:.3cm);
\draw[thick, blue] (-.3,-.6) arc (180:0:.3cm);
}
+
\frac{1}{\phi^{1/2}}
\tikzmath{
\draw[thick, blue] (-.3,.6) arc (-180:0:.3cm);
\draw[thick, blue] (-.3,-.6) arc (180:0:.3cm);
\draw[thick, blue] (0,-.3) -- (0,.3);
\filldraw[blue] (0,-.3) circle (.05cm);
\filldraw[blue] (0,.3) circle (.05cm);
}
\\
\label{eq:FibSkein2}
\tikzmath{
\draw[thick, blue] (0,0) circle (.3cm);
}
&=
\phi
\\
\label{eq:FibSkein3}
\tikzmath{
\draw[thick, blue] (0,0) circle (.3cm);
\filldraw[blue] (0,-.3) circle (.05cm);
\filldraw[blue] (0,.3) circle (.05cm);
\draw[thick, blue] (0,-.6) -- (0,-.3);
\draw[thick, blue] (0,.6) -- (0,.3);
}
&=
\sqrt{\phi}
\cdot
\tikzmath{
\draw[thick, blue] (0,-.6) -- (0,.6);
}
\end{align}
Here, the trivalent vertex represents the normalized basis element $\tau\otimes \tau\to \tau$, which is also rotationally invariant.
We will also need that a triangle evaluates as follows using \eqref{eq:FibSkein1} at the dotted red location below.
\begin{equation}
\label{eq:FibTriangle}
\tikzmath{
\draw[thick, blue] (-.2,-.2) -- (.2,-.2) -- (0,.2) -- (-.2,-.2);
\draw[thick, blue] (-.5,-.5) -- (-.2,-.2);
\draw[thick, blue] (.5,-.5) -- (.2,-.2);
\draw[thick, blue] (0,.2) -- (0,.6);
\filldraw[thick, blue] (0,.2) circle (.05cm);
\filldraw[thick, blue] (-.2,-.2) circle (.05cm);
\filldraw[thick, blue] (.2,-.2) circle (.05cm);
\draw[dotted, red, rounded corners=5pt] (-.35,-.35) rectangle (.35,-.05);
}
=
\sqrt{\phi}
\tikzmath{
\draw[thick, blue] (0,0) circle (.2cm);
\draw[thick, blue] (0,.2) -- (0,.6);
\filldraw[thick, blue] (0,.2) circle (.05cm);
\draw[thick, blue] (-.5,-.6) arc(180:0:.5 and .2);
}
-\frac{1}{\phi^{1/2}}
\tikzmath{
\draw[thick, blue] (-.3,-.3) -- (0,0) -- (.3,-.3);
\draw[thick, blue] (0,0) -- (0,.4);
\filldraw[thick, blue] (0,0) circle (.05cm);
}
=
-\frac{1}{\phi^{1/2}}
\tikzmath{
\draw[thick, blue] (-.3,-.3) -- (0,0) -- (.3,-.3);
\draw[thick, blue] (0,0) -- (0,.4);
\filldraw[thick, blue] (0,0) circle (.05cm);
}
\end{equation}

Now given the Hilbert space vector on the left hand side below which is a diagram drawn on a coral, the gauging map puts the diagram into the punctured skein module on a sphere.

$$
\tikzmath{
\coordinate (a) at (.8,-3);
\coordinate (b11) at (-.5,0);
\coordinate (b21) at (1.5,0);
\coordinate (b12) at (-.25,1.5);
\coordinate (b22) at (1.75,1.5);
\halfDottedEllipse{(a)}{.5}{.2}
\draw[thick] (b12) -- ($ (b12) + (0,-.5) $) .. controls ++(270:.4cm) and ++(90:.4cm) .. ($ (b11) + (.6,0) $) ;
\draw[thick] ($ (b11) + (1,0) $) -- ($ (b11) + (1,-.5) $) .. controls ++(270:.3cm) and ++(110:.3cm) .. ($ (b11) + (1.3,-1.2) $) ;
\draw[thick] ($ (b21) + (1,0) $) -- ($ (b21) + (1,-.5) $) .. controls ++(270:.3cm) and ++(70:.3cm) .. ($ (b21) + (.7,-1.2) $) ;
\draw[thick] (b21) -- ($ (b21) + (0,-.5) $) .. controls ++(270:.3cm) and ++(70:.3cm) .. ($ (b21) + (-.3,-1.2) $) ;
\draw[thick] (a) .. controls ++(90:.8cm) and ++(270:1.2cm) .. ($ (b11) + (0,-.5) $) -- (b11);
\draw[thick] ($ (a) + (1,0) $) .. controls ++(90:.8cm) and ++(270:1.2cm) .. (2.8,-1) -- ($ (b22) + (1,0) $);
\draw[thick] ($ (b12) + (1,0) $) -- ($ (b12) + (1,-.5) $) arc (-180:0:.5) -- (b22);
\draw[thick] (a) arc (-180:0:.5cm);
\foreach \x in {0,2}{
\fill[blue] {(\x,-.7)} circle (.05cm);
\halfDottedEllipse[thick, blue]{(\x-.5,-.5)}{.5}{.2}
}
\halfDottedEllipse[thick, blue]{(0+1.5*2*.125/1.5-.5,1.5-.5)}{.5}{.2}
\draw[thick, blue] ($ (b11) + (.5,-.7) $)  .. controls ++(270:1.5cm) and ++(270:1.5cm) .. ($ (b21) + (.5,-.7) $);
\foreach \x in {0,2}{
\foreach \y in {0,1.5}{
\filldraw[thick, fill=white] (\x+\y*2*.125/1.5,\y) ellipse (.5 and .2);
}}
}
\qquad
\rightsquigarrow
\qquad
\tikzmath{
\foreach \x in {0,1.5}{
\foreach \y in {0,1.5}{
\draw[thick] (\x,\y) circle (.3cm);
}}
\draw[thick,blue] (0,0) circle (.5cm);
\draw[thick,blue] (1.5,0) circle (.5cm);
\draw[thick,blue] (0,1.5) circle (.5cm);
\draw[thick,blue] (.5,0) -- (1,0);
\filldraw[blue] (.5,0) circle (.05cm);
\filldraw[blue] (1,0) circle (.05cm);
\draw[red,dotted,rounded corners=5pt] (-.15,.4) rectangle (.15,1.1);
\draw[red,dotted,rounded corners=5pt] (.3,-.4) rectangle (1.2,-.1);
}
$$
We now apply \eqref{eq:FibSkein1} at the red dotted locations on the right hand side above to get the linear combination
\begin{align*}
&\frac{1}{\phi^2}\cdot
\tikzmath{
\foreach \x in {0,1}{
\foreach \y in {0,1}{
\draw[thick] (\x,\y) circle (.2cm);
}}
\draw[thick, blue, rounded corners=5pt] (-.35,0) -- (-.35,1.35) -- (.35,1.35) -- (.35,.35) -- (1.35,.35) -- (1.35,-.35) -- (-.35,-.35) -- (-.35,0);
\filldraw[blue] (.45,.35) circle (.05cm);
\filldraw[blue] (.75,.35) circle (.05cm);
\draw[thick, blue] (.45,.35) arc(180:0:.15cm);
}
+
\frac{1}{\phi^{3/2}}\cdot
\tikzmath{
\foreach \x in {0,1}{
\foreach \y in {0,1}{
\draw[thick] (\x,\y) circle (.2cm);
}}
\draw[thick, blue, rounded corners=5pt] (-.35,0) -- (-.35,1.35) -- (.35,1.35) -- (.35,.35) -- (1.35,.35) -- (1.35,-.35) -- (-.35,-.35) -- (-.35,0);
\filldraw[blue] (.45,.35) circle (.05cm);
\filldraw[blue] (.75,.35) circle (.05cm);
\draw[thick, blue] (.45,.35) arc(180:0:.15cm);
\draw[thick, blue] (-.35,.65) -- (.35,.65);
\filldraw[blue] (-.35,.65) circle (.05cm);
\filldraw[blue] (.35,.65) circle (.05cm);
}
\\&\qquad
+
\frac{1}{\phi^{3/2}}\cdot
\tikzmath{
\foreach \x in {0,1}{
\foreach \y in {0,1}{
\draw[thick] (\x,\y) circle (.2cm);
}}
\draw[thick, blue, rounded corners=5pt] (-.35,0) -- (-.35,1.35) -- (.35,1.35) -- (.35,.35) -- (1.35,.35) -- (1.35,-.35) -- (-.35,-.35) -- (-.35,0);
\filldraw[blue] (.55,.35) circle (.05cm);
\filldraw[blue] (.75,.35) circle (.05cm);
\draw[thick, blue] (.65,-.35) -- (.65,0);
\draw[thick, blue] (.55,.35) -- (.65,0) -- (.75,.35);
\filldraw[blue] (.65,0) circle (.05cm);
}
+
\frac{1}{\phi}\cdot
\tikzmath{
\foreach \x in {0,1}{
\foreach \y in {0,1}{
\draw[thick] (\x,\y) circle (.2cm);
}}
\draw[thick, blue, rounded corners=5pt] (-.35,0) -- (-.35,1.35) -- (.35,1.35) -- (.35,.35) -- (1.35,.35) -- (1.35,-.35) -- (-.35,-.35) -- (-.35,0);
\filldraw[blue] (.55,.35) circle (.05cm);
\filldraw[blue] (.75,.35) circle (.05cm);
\draw[thick, blue] (.65,-.35) -- (.65,0);
\draw[thick, blue] (.55,.35) -- (.65,0) -- (.75,.35);
\filldraw[blue] (.65,0) circle (.05cm);
\draw[thick, blue] (-.35,.65) -- (.35,.65);
\filldraw[blue] (-.35,.65) circle (.05cm);
\filldraw[blue] (.35,.65) circle (.05cm);
}\,.
\end{align*}
Simplifying with \eqref{eq:FibSkein3} and \eqref{eq:FibTriangle} and expressing the above vectors in the Levin-Wen Hilbert space, 
the first 3 terms simplify as
$$
\frac{1}{\phi^{3/2}}\cdot
\tikzmath{
\draw[dotted, scale=.5] (0,0) grid (2,2);
\draw[thick, blue, scale=.5] (0,0) -- (0,2) -- (1,2) -- (1,1) -- (2,1) -- (2,0) -- (0,0);
}
+
\frac{1}{\phi}\cdot
\tikzmath{
\draw[dotted, scale=.5] (0,0) grid (2,2);
\draw[thick, blue, scale=.5] (0,0) -- (0,2) -- (1,2) -- (1,1) -- (2,1) -- (2,0) -- (0,0);
\draw[thick, blue, scale=.5] (0,1) -- (1,1);
}
-
\frac{1}{\phi^{2}}\cdot
\tikzmath{
\draw[dotted, scale=.5] (0,0) grid (2,2);
\draw[thick, blue, scale=.5] (0,0) -- (0,2) -- (1,2) -- (1,1) -- (2,1) -- (2,0) -- (0,0);
\draw[thick, blue, scale=.5] (1,0) -- (1,1);
}
$$
The last term is more complicated, as the 4-valent space $\mathsf{Fib}(\tau\otimes \tau\to \tau\otimes \tau)$ is 2-dimensional.
For our Levin-Wen model, resolve the 4-valent vertex space as the join of two trivalent vertex spaces as follows:
$$
\tikzmath{
\draw[thick] (-.5,0) -- (.5,0);
\draw[thick] (0,-.5) -- (0,.5);
}
=
\tikzmath{
\draw[thick] (-.5,0) -- (0,0) -- (0,-.5);
\draw[thick] (0,0) -- (.3,.3);
\draw[thick] (.3,.8) -- (.3,.3) -- (.8,.3);
}\,.
$$
In order to express our final diagram in the Levin-Wen space, we compute the skein relation
\begin{align*}
\tikzmath{
\draw[thick, blue] (-.3,0) -- (0,0) -- (0,.3);
\draw[thick, blue] (0,0) -- (.3,-.3);
\draw[thick, blue] (.6,-.3) -- (.3,-.3) -- (.3,-.6);
}
&=
\frac{1}{\phi}
\tikzmath{
\draw[thick, blue] (-.3,0) -- (0,0) -- (0,-.3);
\draw[thick, blue, dotted] (0,0) -- (.3,.3);
\draw[thick, blue] (.3,.9) -- (.3,.3) -- (.9,.3);
\draw[thick, blue] (.3,.6) -- (.6,.3);
}
+
\frac{1}{\phi^{1/2}}
\tikzmath{
\draw[thick, blue] (-.3,0) -- (0,0) -- (0,-.3);
\draw[thick, blue] (0,0) -- (.3,.3);
\draw[thick, blue] (.3,.9) -- (.3,.3) -- (.9,.3);
\draw[thick, blue] (.3,.6) -- (.6,.3);
}
\\&=
\frac{1}{\phi^{1/2}}
\tikzmath{
\draw[thick, blue] (-.3,0) -- (0,0) -- (0,-.3);
\draw[thick, blue, dotted] (0,0) -- (.3,.3);
\draw[thick, blue] (.3,.6) -- (.3,.3) -- (.6,.3);
}
-
\frac{1}{\phi}
\tikzmath{
\draw[thick, blue] (-.3,0) -- (0,0) -- (0,-.3);
\draw[thick, blue] (0,0) -- (.3,.3);
\draw[thick, blue] (.3,.6) -- (.3,.3) -- (.6,.3);
}
\\&=:
\frac{1}{\phi^{1/2}}
\tikzmath{
\draw[thick, blue] (-.3,0) -- (.3,0);
\draw[thick, blue] (0,-.3) -- (0,.3);
\node[blue] at (.1,-.1) {$\scriptstyle 1$};
}
-
\frac{1}{\phi}
\tikzmath{
\draw[thick, blue] (-.3,0) -- (.3,0);
\draw[thick, blue] (0,-.3) -- (0,.3);
\node[blue] at (.1,-.1) {$\scriptstyle \tau$};
}
\end{align*}
This means that the final diagram resolves as
$$
-
\frac{1}{\phi^{3/2}}\cdot
\tikzmath{
\foreach \x in {0,1}{
\foreach \y in {0,1}{
\draw[thick] (\x,\y) circle (.2cm);
}}
\draw[thick, blue, rounded corners=5pt] (-.35,0) -- (-.35,1.35) -- (.35,1.35) -- (.35,.35) -- (1.35,.35) -- (1.35,-.35) -- (-.35,-.35) -- (-.35,0);
\filldraw[blue] (.65,.35) circle (.05cm);
\draw[thick, blue] (.65,-.35) -- (.65,.35);
\draw[thick, blue] (-.35,.65) -- (.35,.65);
\filldraw[blue] (-.35,.65) circle (.05cm);
\filldraw[blue] (.35,.65) circle (.05cm);
}
=
-
\frac{1}{\phi^{2}}\cdot
\tikzmath{
\draw[dotted, scale=.5] (0,0) grid (2,2);
\draw[thick, blue, scale=.5] (0,0) -- (0,2) -- (1,2) -- (1,1) -- (2,1) -- (2,0) -- (0,0);
\draw[thick, blue, scale=.5] (0,1) -- (1,1);
\draw[thick, blue, scale=.5] (1,0) -- (1,1);
\node[blue] at (.6,.4) {$\scriptstyle 1$};
}
+
\frac{1}{\phi^{5/2}}\cdot
\tikzmath{
\draw[dotted, scale=.5] (0,0) grid (2,2);
\draw[thick, blue, scale=.5] (0,0) -- (0,2) -- (1,2) -- (1,1) -- (2,1) -- (2,0) -- (0,0);
\draw[thick, blue, scale=.5] (0,1) -- (1,1);
\draw[thick, blue, scale=.5] (1,0) -- (1,1);
\node[blue] at (.6,.4) {$\scriptstyle \tau$};
}
$$

\section{Conclusion}

In this manuscript, we have shown how to construct the Levin-Wen model as a gauge theory by gauging a new type of trivial theory with a tube algebra $\Tube(\cC)$ symmetry. 
For example, we show how to construct a trivial theory with $\Tube(\mathsf{Fib})$ symmetry. 
This ungauged theory is trivial in the sense that it has a gapped symmetric ground state even when placed on a finite lattice. 
We gauge this theory by considering generalized domain walls and condensing them to create a string-net. 
This construction endows the terms in the Levin-Wen Hamiltonian with a concrete interpretation as measuring gauge theoretic charge and flux.

Our gauging procedure is a twisted gauging procedure in that we start with a trivial theory and produce general string-nets.  
This differs from untwisted gauging where one starts with a non-trivial symmetry protected topological (SPT) phase and produces a twisted quantum double model. 
A direction of future research would be to define a notion of non-trival $\Tube(\cC)$ SPTs which may be gauged in a non-twisted way to obtain general string-nets.

In the case of a group symmetry $G$ where $G$ is finite and solvable, it is known that one can prepare the ground state of the gauge theory in a systematic way. 
This is achieved by sequentially gauging the factor groups of the derived series of $G$ by applying local unitaries and measurements \cite{PRXQuantum.4.020339,PhysRevLett.131.060405,verresen2022efficiently,2112.01519,PhysRevB.108.115144}. 
Another direction of future research would be to explore whether a similar construction exists for $\Tube(\cC)$ symmetries.

\section*{Acknowledgements.}
The authors would like to thank 
Jacob Bridgeman,
Clement Delcamp,
Michael Levin,
Yuan-Ming Lu, and
Nathanan Tantivasadakarn
for helpful discussions.
CJ was supported by NSF DMS 2247202.
KK was supported by NSF DMS 1654159 as well as the
Center for Emergent Materials, an NSF-funded MRSEC, under Grant No. DMR-2011876.
DG, DP, and SS were supported by NSF DMS 2154389.

\bibliographystyle{alpha}
{\footnotesize{
\bibliography{bibliography}
}}
\end{document}